\documentclass[secnumarabic,aps,pra,amsfonts,twocolumn,%
showkeys,floatfix%
]{revtex4-1}

\usepackage{bm,amsfonts,amsthm,amsmath}
\usepackage{placeins}
\usepackage{graphicx}
\newcommand{\refg}[1]{(\ref{#1})}

\newtheorem{thm}{Theorem}
\newtheorem{definition}{Definition}

\begin{document}
\title{Group theory of Wannier functions providing the basis for a
  deeper understanding of magnetism and superconductivity}
\author{Ekkehard Kr\"uger}\author{Horst P. Strunk}
\affiliation{Institut f\"ur Materialwissenschaft, Materialphysik,
  Universit\"at Stuttgart, D-70569 Stuttgart, Germany}
%
\date{\today}
\begin{abstract}
  The paper presents the group theory of best localized and
  symmetry-adapted Wannier functions in a crystal of any given space
  group G or magnetic group M. Provided that the calculated band
  structure of the considered material is given and that the symmetry
  of the Bloch functions at all the points of symmetry in the
  Brillouin zone is known, the paper details whether or not the Bloch
  functions of particular energy bands can be unitarily transformed
  into best localized Wannier functions symmetry-adapted to the space
  group G, to the magnetic group M, or to a subgroup of G or M. In
  this context, the paper considers usual as well as spin-dependent
  Wannier functions, the latter representing the most general
  definition of Wannier functions. The presented group theory is a
  review of the theory published by one of the authors in several
  former papers and is independent of any physical model of magnetism
  or superconductivity. However, it is suggested to interpret the
  special symmetry of the best localized Wannier functions in the
  framework of a nonadiabatic extension of the Heisenberg model, the
  nonadiabatic Heisenberg model. On the basis of the symmetry of the
  Wannier functions, this model of strongly correlated localized
  electrons makes clear predictions whether or not the system can
  possess superconducting or magnetic eigenstates.
\end{abstract}

\keywords{Wannier functions, magnetism, superconductivity, group theory}
\maketitle

\section{Introduction}
The picture of strongly correlated localized or nearly-localized
electrons is the base of a successful theoretical description of both
high-temperature superconductivity and magnetism (see, e.g.,
~\cite{scalapino,lechermann, metzner} and citations given there). In
almost all cases the appertaining localized electron states are
represented by atomic orbitals that define, for instance, partially
filled $s$-, $d$-, or $p$- bands.

Another option would be to represent the localized electron states by
best localized and symmetry-adapted Wannier functions. In contrast to
atomic functions, Wannier functions situated at {\em adjacent} atoms
are {\em orthogonal} and, hence, electrons occupying (temporarily)
adjacent localized states represented by Wannier functions comply with
the Pauli principle. In addition, Wannier function form a {\em
  complete} set of basis functions within the considered narrow,
partially filled band. Consequently, Wannier functions contain {\em
  all} the physical information about this energy band.

Wannier functions tend to be ignored by the theory of
superconductivity and magnetism because we need a {\em closed} complex
of energy bands [Definition~\ref{def:2}] for the construction of {\em
  best localized} Wannier functions.  Such closed complexes, however,
do not exist in the band structures of the metals where all the bands
are connected to each other by band degeneracies.

Fortunately, this problem can be solved in a natural way by
constructing Wannier functions with the reduced symmetry of a magnetic
group or by constructing spin-dependent Wannier functions as shall be
details in the present paper. In both cases, interfering
band degeneracies are sometimes removed in the band structure with
the reduced symmetry.

Against the background of the described characteristics of the Wannier
functions, our following two observations should not be too
surprising:
\begin{enumerate}
\item Materials possessing a magnetic structure with the magnetic
  group $M$ also possess a closed, narrow and roughly half-filled
  complex of energy bands in their band structure whose Bloch
  functions can be unitarily transformed into best localized Wannier
  functions that are symmetry-adapted to the magnetic group $M$. These
  energy bands form a ``magnetic band'', see Definition~\ref{def:21}.
\item Both normal and high-temperature superconductors (and only
  superconductors) possess a closed, narrow and roughly half-filled
  complex of energy bands in their band structure whose Bloch spinors
  can be unitarily transformed into best localized {\em spin-dependent}
  Wannier functions that are symmetry-adapted to the (full) space
  group $G$ of the material. These energy bands form a
  ``superconducting band'', see Definition~\ref{def:22}.
\end{enumerate}

The first observation (i) was made at the band structures of
Cr~\cite{ea}, Fe~\cite{ef}, La$_2$CuO$_4$~\cite{josla2cuo4},
YBa$_2$Cu$_3$O$_6$~\cite{ybacuo6}, undoped LaFeAsO~\cite{lafeaso1},
and BaFe$_2$As$_2$~\cite{bafe2as2}; the second observation (ii) at the
band structures of numerous elemental superconductors~\cite{es2} and
of the (high-temperature) superconductors
La$_2$CuO$_4$~\cite{josla2cuo4}, YBa$_2$Cu$_3$O$_7$~\cite{josybacuo7},
MgB$_2$~\cite{josybacuo7}, and doped LaFeAsO~\cite{lafeaso2}. It is
particularly important that partly filled superconducting bands cannot
be found in those elemental metals (such as Li, Na, K, Rb, Cs, Ca Cu,
Ag, and Au) which do not become superconducting~\cite{es2}. An
investigation into the band structures of the transition metals in
terms of superconducting bands straightforwardly leads to the Matthias
rule~\cite{josm}.

Though these two observations are clear, their theoretical
interpretation is initially difficult. This is primarily due to the
fact that the models of localized electrons developed so far, as,
e.g., the familiar Hubbard model~\cite{hubbard}, are tailored to {\em
  atomic orbitals} that represent the localized states during an
electronic hopping motion. Within modern theoretical concepts, the
Wannier functions often are nothing but a complete basis in the space
spanned by the Bloch functions. Thus, their symmetry is often believed
to do not tell anything about the physics of strongly correlated
electrons.

In the light of this background, we suggest to interpret the special
symmetries of best localized Wannier functions within the nonadiabatic
Heisenberg model~\cite{enhm,josybacuo7}. This model of strongly
correlated localized electrons starts in a consistent way from
symmetry-adapted and best localized {\em Wannier} functions that
represent the localized electron states related to the hopping motion
and defines the Hamiltonian $H^n$ of the related nonadiabatic system.
On the basis of the symmetry of the Wannier functions, the
nonadiabatic model makes clear predictions whether or not $H^n$ can
possess superconducting or magnetic {\em
  eigenstates}~\cite{es2,josn,ea,bafe2as2}. In this context, the
nonadiabatic Heisenberg model no longer uses terms like $s$-, $p$-, or
$d$-bands, but only speaks of superconducting or magnetic bands.

In particularly interesting cases, the nonadiabatic Heisenberg model
predicts that a small distortion of the lattice or a doping is
required for the stability of the superconducting or magnetic
state. Thus, in undoped LaFeAsO~\cite{lafeaso1} and in
BaFe$_2$As$_2$~\cite{bafe2as2} the antiferromagnetic state must be
stabilized by an experimentally well established
distortion~\cite{huang2,clarina}, while in
YBa$_2$Cu$_3$O$_6$~\cite{ybacuo6} it is stable in the undistorted
crystal. Superconductivity in LaFeAsO~\cite{lafeaso1} requires the
experimentally confirmed doping~\cite{clarina,
  nomura,kitao,nakai}. Also the superconducting state in
LiFeAs~\cite{lifeas} should be accompanied by a small distortion of
the lattice which, to our knowledge, is experimentally not yet
confirmed. Superconductivity in YBa$_2$Cu$_3$O$_7$~\cite{josybacuo7},
MgB$_2$~\cite{josybacuo7} as well as in the transition
elements~\cite{es2} (such as in Nb~\cite{josn}), on the other hand,
does not require any distortion or doping.

In the case of (conventional and high-$T_c$~\cite{ehtc})
superconductivity, the nonadiabatic Heisenberg model provides a new
mechanism of Cooper pair formation which may be described in terms of
{\em constraining forces}~\cite{josn} and {\em spring-mounted Cooper
  pairs}~\cite{josi}.

Any application of the nonadiabatic Heisenberg model starts with a
determination of the symmetry of best localized (spin-dependent)
Wannier functions related to the band structure of the material under
consideration.  In the following we shall summarize and update the
group theory of Wannier functions as published in former papers and
give a detailed description how to determine the symmetry of best
localized Wannier functions if they exist in the given band
structure. Though we shall also define the two terms ``magnetic'' and
``superconducting'' bands which are related to the nonadiabatic
Heisenberg model, the presented group theory {\em is independent of
  any physical model of magnetism or superconductivity}.

\setcounter{equation}{0}
\setcounter{definition}{0}
\setcounter{thm}{0}
\section{Usual (spin-independent) Wannier functions}
\subsection{Definition}
\label{sec:defwf}
Consider a closed complex of $\mu$ energy bands in the band structure
of a metal or a semiconductor.
\begin{definition}[closed]
\label{def:2}
A complex of energy bands is called closed if the bands are not
connected by degeneracies to bands not belonging to the complex.
\end{definition}
\begin{definition}[closed band]
\label{def:14}
In the following a closed complex of $\mu$ energy bands is referred to
as a single closed band consisting of $\mu$ branches.
\end{definition}

The metals do not possess closed bands in their band
structures. However, closed bands may arise after the activation of a
perturbation reducing the symmetry in such a way that interfering band
degeneracies are removed. Such a reduction of the symmetry may be
caused by a magnetic structure or by a (slight) distortion of the
crystal.

Hence, we assume that the Hamiltonian $\mathcal{H}$ of a single
electron in the considered material consists of a part $\mathcal{H}_G$
with the unperturbed space group $G$ and a perturbation
$\mathcal{H}_H$ with the space group $H$,
\begin{equation}
  \label{eq:93}
  \mathcal{H} = \mathcal{H}_G + \mathcal{H}_H,  
\end{equation}
where $H$ is a subgroup of $G$,
\begin{equation}
  \label{eq:26}
  H \subset G.
\end{equation}
In general, the considered closed energy band of $\mu$ branches was
not closed before the perturbation $\mathcal{H}_H$ was activated.

Assume the Bloch functions $\varphi_{\bm k,q}(\bm r)$ (labeled by the
wave vector $\bm k$ and the branch index $q$) as the solutions of the
Schr\"odinger equation of $\mathcal{H}$ to be completely calculated in
the first domain of the Brillouin zone.

\begin{definition}[first domain]
\label{def:1}
Let be $h$ the order of the point group $H_0$ of $H$. Then the
Brillouin zone is divided by the planes of symmetry into $h$
domains. An arbitrary chosen domain we call the first domain. This
first domain shall comprise the bounding planes, lines and points of
symmetry, too.
\end{definition}

As in Ref.~\cite{ew1}, in the rest of the Brillouin zone the Bloch
functions shall be determined by the equation
\begin{equation}
  \label{eq:95}
  \varphi_{\alpha \bm k,q}(\bm r) = P(\{\alpha | \bm t_{\alpha }\})\varphi_{\bm
  k,q}(\bm r)\text{ for } \alpha \in H_0,
\end{equation}
where $\bm k$ lies in the first domain, and in the $\bm k$ space
outside the Brillouin zone by the equation
\begin{equation}
  \label{eq:96}
  \varphi_{\bm k + \bm K,q}(\bm r) = \varphi_{\bm k,q}(\bm r).
\end{equation}
$\bm K$ denotes a vector of the reciprocal lattice and $H_0$ stands
for the point group of $H$.

\begin{definition}[symmetry operators]
\label{def:9}
$P(a)$ denotes the symmetry operator assigned to the space group
operation $a = \{\alpha|\bm t_{\alpha}\}$ consisting of a point group
operation $\alpha$ and the associated translation $\bm t_{\alpha}$,
acting on a wave function $f(\bm r)$ according to
\begin{equation}
  \label{eq:69}
P(a)f(\bm r) = f(a^{-1}\bm r) = f(\alpha^{-1}\bm r - \alpha^{-1}\bm t_{\alpha}).
\end{equation}
\end{definition}

The Bloch functions $\varphi_{\bm k,q}(\bm r)$ of the closed band
under observation can be unitarily transformed into Wannier functions
\begin{equation}
  \label{eq:1}
w_i(\bm r - \bm R - \bm\rho_i) = \frac{1}{\sqrt{N}}\sum^{BZ}_{\bm
  k}e^{-i\bm k (\bm R + \bm\rho_i)}\widetilde\varphi_{\bm k,i}(\bm r)   
\end{equation}
centered at the positions $\bm R + \bm\rho_i$, 
where the functions
\begin{equation}
  \label{eq:2}
\widetilde\varphi_{\bm k,i}(\bm r) = \sum_{q = 1}^{\mu} g_{iq}(\bm k
  )\varphi_{\bm k,q}(\bm r)   
\end{equation}
are ``generalized'' Bloch functions~\cite{ew1}. 
 The sum in Eq.~\refg{eq:1} runs
over the $N$ vectors $\bm k$ of the first Brillouin zone (BZ), the sum
in Eq.~\refg{eq:2} over the $\mu$ branches of the considered
band, $\bm R$ denote the vectors of the Bravais lattice, and the
coefficients $g_{iq}(\bm k)$ in Eq.~\refg{eq:2} are the elements of an
unitary matrix ${\bf g}(\bm k)$,
\begin{equation}
  \label{eq:5}
{\bf g}^{-1}(\bm k) = {\bf g}^{\dagger}(\bm k),
\end{equation}
in order that the Wannier functions are orthonormal,
\begin{equation}
  \label{eq:52}
  \int w^*_i(\bm r - \bm R - \bm\rho_i)w_{i'}(\bm r - \bm R' -
  \bm\rho_{i'})d\bm r
  = \delta_{\bm R\bm R'}\delta_{ii'}.
\end{equation}

\begin{definition}[best localized]
\label{def:3}
The Wannier functions are called best localized if the
coefficients $g_{iq}(\bm k)$ may be chosen in such a way that the
generalized Bloch functions $\widetilde\varphi_{\bm k,i}(\bm r)$ move
-- for fixed $\bm r$ -- {\em continuously} through to whole $\bm k$ space~\cite{ew1}.
\end{definition}

As it was already shown in Ref.~\cite{bouckaert}, the Bloch functions
$\varphi_{\bm k,q}(\bm r)$ as the eigenfunctions of the Hamiltonian
$\mathcal{H}$ may be chosen in such a way that they vary continuously
as functions of $\bm k$ through the first domain and, in particular,
approach continuously the boundaries of the first domain. From
Eqs.~\refg{eq:95} and~\refg{eq:96}, however, we {\em cannot} conclude
that they also {\em cross} continuously the boundaries of the domains within
the Brillouin zone or at the surface of the Brillouin
zone. Fortunately, this problem is solvable by group-theoretical
methods~\cite{ew1,ew2}. Theorem~\ref{theorem1} shall define the
condition for best localized and symmetry-adapted
[Definition~\refg{def:15}] Wannier functions.

\subsection{Symmetry-adapted Wannier functions}
\label{sec:sawf}
In Ref.~\cite{ew1} we demanded that symmetry-adapted Wannier functions
satisfy the equation
\begin{equation}
  \label{eq:6}
w_i\big(\alpha^{-1}(\bm r - \bm R - \bm\rho_i)\big) = \sum_{j =
  1}^{\mu}D_{ji}(\alpha )w_j(\bm r - \bm R - \bm\rho_i) 
\end{equation}
for the elements $\alpha$ of the point group $H_0$ of $H$,
where the $D_{ji}(\alpha )$ are the elements of the matrices
\begin{equation} 
  \label{eq:20}
  \text{\bf D}(\alpha ) = \big[D_{ij}(\alpha )\big]  
\end{equation}
forming a representation $\bm D$ of $H_0$ which in most cases is
reducible. [It should be noted that the sum in Eq.~\refg{eq:6} runs
over $w_j(\bm r - \bm R - \bm\rho_i)$ and not over $w_j(\bm r - \bm R
- \bm\rho_j)$.]

Eq.~\refg{eq:6} defines the symmetry of Wannier functions in general
terms, particularly they may be centered at a variety of positions
$\bm \rho_i$ being different from the positions of the atoms. However,
in the context of superconducting and magnetic bands we may restrict
ourselves to Wannier functions centered at the positions of the atoms.

Thus, we assume 
\begin{enumerate}
\item that the positions $\bm\rho_i$ of the Wannier functions in
  Eq.~\refg{eq:1} are the positions of atoms,
\item that only atoms of the same sort are considered (although, of
  course, other atoms may exist), and
\item that there is one Wannier function at each atom.
\end{enumerate}

Under these assumptions~\cite{enhm}, 
\begin{itemize}
\item
the Wannier functions may be
labeled by the positions of the atoms,
\begin{equation}
\label{eq:7}
w_{\bm T}(\bm r) \equiv w_i(\bm r - \bm R - \bm\rho_i),  
\end{equation}
where
\begin{equation}
  \label{eq:22}
\bm T = \bm R + \bm\rho_{i},
\end{equation}
\item the matrix representatives $\text{\bf D}(\alpha )$ of the
  representation $\bm D$ in Eq.~\refg{eq:6} have {\em one}
  non-vanishing element $D_{ij}(\alpha )$ with
\begin{equation}
  \label{eq:21}
  |D_{ij}(\alpha )| = 1
\end{equation}
in each row and each column, and 
\item
Eq.~\refg{eq:6} may be written in the
considerably simplified form
\begin{equation}
  \label{eq:8}
P(a)w_{\bm T}(\bm r) = D_{ji}(\alpha )w_{\bm T'}(\bm r)
\mbox{ for } a \in H
\end{equation}
where
\begin{equation}
  \label{eq:9}
\bm T' = \alpha\bm T + \bm t_{\alpha}
\end{equation}
and the subscripts $i$ and $j$ denote the number of the atoms at
position $\bm T$ and $\bm T'$, respectively.  
\end{itemize}
\begin{definition}[number of the atom]
\label{def:7}
The subscript $i$ of the vector $\bm \rho_i$ in Eq.~\refg{eq:22}
defines the number of the atom at position $\bm T$.
\end{definition}

\begin{definition}[symmetry-adapted]
\label{def:15}
We call the Wannier functions symmetry-adapted to $H$ if they
satisfy Eq.~\refg{eq:8}.
\end{definition}

\begin{thm}
\label{theorem10}
The third assumption (iii) shows immediately that the number $\mu$ of
the branches of the band under observation equals the number of the
considered atoms in the unit cell.
\end{thm}

Eqs.~\refg{eq:8} and~\refg{eq:9} {\em define}
the non-vanishing elements and, hence, we may write Eq.~\refg{eq:21} more
precisely,
\begin{equation}
  \label{eq:43}
  |D_{ji}(\alpha )| = 
\left\{
\begin{array}{ll}
1 & \text{if } \alpha\bm \rho_i + \bm t_{\alpha} = \bm \rho_j  + \bm R \\
0 & \text{else,}\\
\end{array}
\right.
\end{equation}
where $\{\alpha | \bm t_{\alpha}\} \in H$ and $\bm R$ still denotes a lattice
vector. 

\begin{definition}[the representation defining the Wannier functions]
\label{def:4}
In what follows, the representation $\bm D$ of $H_0$ with the matrix
representatives $${\bf D}(\alpha ) = [D_{ij}(\alpha )]$$ defined by
Eq.~\refg{eq:8} shall be shortly referred to as ``the representation
defining the Wannier functions'' and its matrix representatives ${\bf
  D}(\alpha )$ to as ``the matrices defining the Wannier functions''.
\end{definition}

\begin{definition}[unitary generalized permutation matrices]
\label{def:16}
Since the matrices $\text{\bf D}(\alpha )$ defining the Wannier
functions have {\em one} non-vanishing element obeying
Eq.~\refg{eq:43} in each row and each column, they are so-called
unitary generalized permutation matrices.
\end{definition}

\setcounter{equation}{0}
\setcounter{definition}{0}
\setcounter{thm}{0}

\section{Determination of the representations $\bm D$ defining the
  Wannier functions}
\label{sec:matrices}
In the following Sec~\ref{sec:condition} we shall give a simple
condition [Theorem~\ref{theorem1}] for best localized and
symmetry-adapted Wannier functions yielding the representations of the
Bloch functions at all the points $\bm k$ of symmetry in the Brillouin
zone. However, in Theorem~\ref{theorem1} the representations $\bm D$
defining the Wannier functions must be known. Hence, first of all we
have to determine in this section all the possible representations
that may define the Wannier functions. In this context we assume
first that all the atoms are connected by symmetry. This restricting
assumption shall be abandoned not until in
Sec.~\ref{sec:atomsnotconnected}.
\begin{definition}[connected by symmetry]
\label{def:8}
Two atoms at
positions $\bm \rho_i$ and $\bm \rho_j$ are connected by symmetry
if there exists at least one element $a = \{\alpha |\bm t_{\alpha}\}$
in the space group $H$ satisfying the equation
\begin{equation}
  \label{eq:16}
  \alpha\bm \rho_i + \bm t_{\alpha} = \bm \rho_j  + \bm R,   
\end{equation}
where $\bm R$ is a lattice vector. 
\end{definition}

\subsection{General properties of the representatives $\text{\bf
    D}(\alpha )$ of $\bm D$}
\label{sec:general}
First consider the diagonal elements
\begin{equation}
  \label{eq:74}
  d_i(\alpha ) = D_{ii}(\alpha )
\end{equation}
of the matrices $\text{\bf D}(\alpha )$ defining the Wannier
functions. From Eq.~\refg{eq:43} we obtain
\begin{equation}
  \label{eq:12}
  |d_i(\alpha )| = 
\left\{
\begin{array}{ll}
1 & \text{if } \alpha\bm \rho_{i} + \bm t_{\alpha} = \bm \rho_{i} + \bm R \\
0 & \text{else}\\
\end{array}
\right.
\end{equation}
where $\bm R$ denotes a lattice vector. This equation demonstrates
that the matrix $\text{\bf D}(\alpha )$ has non-vanishing diagonal
elements $d_i(\alpha )$ if the space group operation $a = \{\alpha|\bm
t_{\alpha}\}$ leaves invariant the position $\bm \rho_{i}$ of the
$i$th atom. These space group operations form a group, namely the
group $G_{\bm \rho_i}$ of the position $\bm \rho_i$.
\begin{definition}[group of position]
  \label{def:10}
The group $G_{\bm \rho_i}$ of the position $\bm \rho_i$ is defined by
\begin{equation}
  \label{eq:11}
  a \in G_{\bm \rho_i} \text{ if } a \in H \text{ and
  } \alpha\bm \rho_{i} + \bm t_{\alpha} = \bm \rho_{i} + \bm R.
\end{equation}  
$G_{0\bm \rho_i}$ denotes the point group of $G_{\bm \rho_i}$.
\end{definition}

Hence, the non-vanishing diagonal elements $d_i(\alpha )$ of the
matrices $\text{\bf D}(\alpha )$ form a one-dimensional representation
$\bm d_i$ of the point group $G_{0\bm \rho_i}$ of $G_{\bm
  \rho_i}$. The Wannier functions transform according to
\begin{equation}
  \label{eq:13}
  P(a)w_{\bm T}(\bm r) = d_i(\alpha )w_{\bm T + \bm R}(\bm r) \text{ for } \alpha \in 
  G_{0\bm \rho_i}
\end{equation}
[cf. Eq.~\refg{eq:8}] by application of a space group operator
$P(a)$ (where $\bm R$ still denotes a vector of the Bravais lattice). 
From Eq.~\refg{eq:6} we may derive the equivalent equation
\begin{equation}
  \label{eq:14}
  w_i\big(\alpha^{-1}(\bm r - \bm R - \bm\rho_i)\big)  = 
  d_i(\alpha )w_i(\bm r - \bm R - \bm\rho_i)\text{ for } \alpha \in 
  G_{0\bm \rho_i}
\end{equation}
or, after shifting the origin of the coordinate system into the center
of the function $w_i(\bm r - \bm R - \bm\rho_i)$,
$$\bm r' = \bm r - \bm R - \bm\rho_i, $$
we receive an equation
\begin{equation}
  \label{eq:40}
  w_i(\alpha^{-1}\bm r')  = 
  d_i(\alpha )w_i(\bm r')\text{ for } \alpha \in 
  G_{0\bm \rho_i}
\end{equation}
emphasizing the point-group symmetry of the
Wannier function at position $\bm R + \bm\rho_i$.

In constructing the representation $\bm D$ defining the Wannier
functions we cannot arbitrarily chose the one-dimensional
representations ${\bm d_i}$ of $G_{0\bm \rho_i}$ because they must
be chosen in such a way that the matrix representatives $\text{\bf
  D}(\alpha )$ form a representation of the point group $H_0$, i.e.,
they must obey the multiplication rule
\begin{equation}
  \label{eq:28}
  \text{\bf D}(\alpha\beta ) = \text{\bf D}(\alpha )\text{\bf D}(\beta ) 
\end{equation}
for all the elements $\alpha$ and $\beta$ in $H_0$.  

In what follows we assume that all the groups $G_{\bm \rho_i}$ are
{\em normal} subgroups of $H$. In fact, in all the crystal structures
we examined in the past, $G_{\bm \rho_i}$ was a normal subgroup, be it
because it was a subgroup of index 2 or be it because it was the
intersection of two subgroups of index 2. Both cases are sufficient
for a normal subgroup. We believe that in all physically relevant
crystal structures $G_{\bm \rho_i}$ is a normal subgroup of $H$. If
not, the present formalism must be extended for these structures.

When the groups $G_{\bm \rho_i}$ are normal subgroups of $H$, each of
the groups $G_{\bm \rho_i}$ contains only {\em complete} classes of
$H$,
\begin{equation}
  \label{eq:44}
  b^{-1}ab \in G_{\bm \rho_i} \text{ if } a \in G_{\bm \rho_i} \text{ and } b \in H.
\end{equation}
We now show that, as a consequence, all the groups $G_{\bm \rho_i}$
contain the {\em same} space group operations.  

Let be $b = \{\beta | \bm t_{\beta}\}$ a space group operation of $H$
moving $\rho_i$ into $\rho_j$,
$$\beta\bm \rho_i + \bm t_{\beta} = \bm \rho_j + \bm R,$$ 
then 
\begin{equation}
  \label{eq:47}
  c = b^{-1}ab
\end{equation}
is an element of $G_{\bm \rho_i}$ if $a \in G_{\bm
  \rho_j}$. Eq.~\refg{eq:47} even yields {\em all} the elements $c$ of
$G_{\bm \rho_i}$ when $a$ runs throw all the elements of $G_{\bm
  \rho_j}$ because we may write Eq.~\refg{eq:47} in the form   
$$ bcb^{-1} = a $$
showing that we may determine from {\em any} element $c \in G_{\bm
  \rho_i}$ an element $a \in G_{\bm \rho_j}$.

On the other hand, Eq.~\refg{eq:44} shows that $c$ is an element of
$G_{\bm \rho_j}$, too. When $a$ runs through all the elements of
$G_{\bm \rho_j}$, then also $c$ runs through all the elements of
$G_{\bm \rho_j}$. Consequently, all the groups $G_{\bm \rho_i}$ as well as
all the related point groups $G_{0\bm \rho_i}$ contain the same
elements.

Thus, we may omit the index $i$ and define
\begin{definition}[group of position]
\label{def:11}
The group $G_p$ and the related point group $G_{0p}$ of the positions
of the atoms is defined by
\begin{equation}
  \label{eq:45}
  G_p \equiv G_{\bm \rho_i}
\end{equation}
and
\begin{equation}
  \label{eq:46}
  G_{0p} \equiv G_{0\bm \rho_i},
\end{equation}
respectively, where $G_{\bm \rho_i}$ and $G_{0\bm \rho_i}$ are given by
Definition~\ref{def:10}.
\end{definition}

\subsection{Necessary condition for of the representatives $\text{\bf
    D}(\alpha )$ of $\bm D$} 
\label{sec:necessary}

The one-dimensional representations $\bm d_i$ of $G_{0p}$ must be
chosen in such a way that the matrices ${\bf D}(\alpha )$ defining the
Wannier functions form a representation $\bm D$ of the {\em complete}
point group $H_0$. A necessary condition is given by the evident
Theorem~\ref{theorem2}.

\begin{thm}
\label{theorem2}
If the matrices ${\bf D}(\alpha )$ {\em cannot} be completely reduced
into the irreducible representations of $H_0$, then they do not form a
representation of the point group $H_0$.
\end{thm}
This theorem is necessary, but not sufficient: even if the matrices
${\bf D}(\alpha )$ can be completely reduced into the irreducible
representations of $H_0$ then they need not form a representation of
the point group $H_0$~\cite{streitwolf}. The complete decomposition of
a reducible representation is described, e.g., in
Refs.~\cite{streitwolf} and~\cite{bc}, in particular, see Eq.~(1.3.18)
of Ref.~\cite{bc}.
Theorem~\ref{theorem2} leads to three important cases:

\begin{itemize}
\item Case (i): If all the representations $\bm d_i$ are subduced from
  {\em one}-dimensional representations of $H_0$, then all the
  representations $\bm d_i$ are equal,
  \begin{equation}
    \label{eq:18}
    \bm d_i = \bm d \quad\text{for all the positions } \bm
      \rho_i.
  \end{equation}
  The representation ${\bm d}$ may be equal to any one-dimensional
  representation of $G_{0p}$ subduced from a {\em one}-dimensional
  representation of $H_0$.
\item Case (ii): If all the representations $\bm d_i$ are subduced from
  {\em two}-dimensional representations of $H_0$, then one half of the
  representations $\bm d_i$ is equal to $\bm d_A$ and the other
  half is equal to $\bm d_B$,
  \begin{equation}
    \label{eq:19}
    \begin{array}{lll}
      \bm d_i & = & \bm d_A\quad\text{for one half of the
        positions } \bm\rho_i\\
      \bm d_i & = & \bm d_B\quad\text{for the remaining positions } \bm \rho_i,\\ 
    \end{array}
  \end{equation}
  where $\bm d_A$ and $\bm d_B$ are subduced from the {\em
    same} two-dimensional representation of $H_0$. In special cases,
  the two representations $\bm d_A$ and $\bm d_B$ may be
  equal, see below.
\item Case (iii): ``Mixed'' representations $\bm D$ consisting of both
  representations $\bm d_i$ subduced from one- and two-dimensional
  representations of $H_0$ do not exist.
\end{itemize}
A further case that the representations $\bm d_i$ are subduced from
{\em three}-dimensional representations of $H_0$ may occur in crystals
of high symmetry but is not considered in this paper.

These results (i) -- (iii) follow from the very fact that
Eq.~\refg{eq:8} describes an interchange of the Wannier functions at
different positions $\bm\rho_i$. Such an interchange, however, does
not alter the symmetry of the Wannier functions.

\subsection{Sufficient condition for of the representatives $\text{\bf
    D}(\alpha )$ of $\bm D$}
\label{sec:sufficient}

For $\alpha \in G_{0p}$ the matrices ${\bf D}(\alpha )$ defining the
Wannier functions are diagonal, while the remaining matrices ${\bf
  D}(\alpha )$ [for $\alpha \in H_0 - G_{0p}$] do not possess any
diagonal element. Theorem~\ref{theorem2} only gives information about
the diagonal matrices ${\bf D}(\alpha )$. Hence, this theorem indeed
cannot be sufficient because we do not know whether or not the
remaining matrices obey the multiplication rule~\refg{eq:28}.

In this section we assume that the matrices ${\bf D}(\alpha )$ already
satisfy Theorem~\ref{theorem2} and examine the conditions under which
they actually form a (generally reducible) representation of
$H_0$. In doing so, we consider separately the two cases (i) and (ii)
of the preceding Sec.~\ref{sec:necessary}.

\subsubsection{Case (i) of Sec.~\ref{sec:necessary}}
\label{sec:onedimensional}
No further problems arises when case (i) of Sec.~\ref{sec:necessary}
is realized. In this case, Theorem~\ref{theorem2} is necessary {\em
  and} sufficient. To justify this assertion, we write down explicitly
the non-diagonal elements of the matrices ${\bf D}(\alpha )$.

Let be $\bm \delta$ any one-dimensional representation of $H_0$
subducing the representation $\bm d$ in Eq.~\refg{eq:18}. If we
put all the non-vanishing elements of the matrices ${\bf
  D}(\alpha )$ equal to the elements $\delta(\alpha )$ of $\bm
\delta$,
\begin{equation}
  \label{eq:42}
  D_{ji}(\alpha ) = 
\left\{
\begin{array}{ll}
\delta(\alpha ) & \text{ if } \alpha\bm \rho_i + \bm t_{\alpha} = \bm \rho_j  + \bm R \\
0 & \text{ else,}\\
\end{array}
\right.
\end{equation}
then we receive matrices ${\bf D}(\alpha )$ evidently
multiplying as the elements of the representation $\bm \delta$ and,
consequently, obeying the multiplication rule in Eq.~\refg{eq:28}.

\subsubsection{Case (ii) of Sec.~\ref{sec:necessary}}
\label{sec:twodimensional}
The situation is a little more complicated when case (ii) of
Sec.~\ref{sec:necessary} is realized. Now, the representations $\bm
d_A$ and $\bm d_B$ in Eq.~\refg{eq:19} may be distributed across the
positions $\bm \rho_i$ in such a way that the matrices ${\bf D}(\alpha
)$ form a representation of $H_0$ or do not. Though we always find a
special distribution of the $\bm d_A$ and $\bm d_B$ yielding matrices
${\bf D}(\alpha )$ actually forming a representation of $H_0$, we have
to rule out those distributions not leading to a representation of
$H_0$, because in the following [in Eqs.~\refg{eq:25},~\refg{eq:61},
and~\refg{eq:116}] we need the matrices ${\bf D}(\alpha )$ explicitly.

Let be $\bm \Delta$ [with the matrix representatives $\bm
\Delta(\alpha )$] a two-dimensional representation of $H_0$ subducing
the two representations $\bm d_A$ and $\bm d_B$ of $G_{0p}$. The
matrix representatives $\bm \Delta(\alpha )$ may be determined, e.g.,
from Table 5.1 of Ref.~\cite{bc}.

As a first step, $\bm \Delta$ must be unitarily transformed (by a
matrix {\bf Q}) in such a way that the matrices $\bm \Delta(\alpha )$
are diagonal for $\alpha \in G_{0p}$,
\begin{equation}
  \label{eq:48}
  \overline{\bm \Delta}(\alpha ) 
   = {\bf Q}^{-1}{\bm \Delta}(\alpha ){\bf Q}\quad = \text{ diagonal for }\alpha \in G_{0p}. 
\end{equation}

Now consider a certain distribution of the representations $\bm d_A$
and $\bm d_B$ across the positions $\bm \rho_i$.  Then we may
determine the elements of the matrices ${\bf D}(\alpha )$, if they
exist, be means of the formula
\begin{equation}
  \label{eq:49}
\begin{array}{l}
\text{if}\quad\alpha\bm \rho_i + \bm t_{\alpha} = \bm \rho_j  + \bm R\\
\\
D_{ji}(\alpha ) = \left\{
\begin{array}{lll}
  \overline{\Delta}_{12}(\alpha ) & \text{ if } \bm d_j = \bm
    d_A  & \text{and }\bm d_i = \bm d_B,\\
\\
  \overline{\Delta}_{21}(\alpha ) & \text{ if } \bm d_j = \bm
    d_B  & \text{and }\bm d_i = \bm d_A,\\
\\
  \overline{\Delta}_{11}(\alpha ) & \text{ if } \bm d_j = \bm
    d_A  & \text{and }\bm d_i = \bm d_A,\\
\\
  \overline{\Delta}_{22}(\alpha ) & \text{ if } \bm d_j = \bm
    d_B  & \text{and }\bm d_i = \bm d_B,\\
\end{array}
\right.\\
\\
\text{else} \\
\\
D_{ji}(\alpha ) = 0,\\
\\
\end{array}
\end{equation}\\
where the $\overline{\Delta}_{ij}(\alpha )$ denote the elements of
$\overline{\bm \Delta}(\alpha )$.

It turns out that in each case the matrices determined by
Eq.~\refg{eq:49} satisfy the multiplication rule in Eq.~\refg{eq:28}
if Eq.~\refg{eq:49} produces for each space group operation $a \in H$
an unitary generalized permutation matrix ${\bf D}(\alpha )$. This may
be understood because Eq.~\refg{eq:49} defines the complex numbers
$D_{ji}(\alpha )$ in such a way that the Wannier functions transform
in Eq.~\refg{eq:8} in an unequivocal manner like the basis functions
for $\overline{\bm \Delta}$. With ``like'' the basis functions we want
to express that by application of any space group operator $P(\{\alpha
|\bm t_{\alpha }\})$ they are multiplied in Eq.~\refg{eq:8} by the
same complex number $\overline{\Delta}_{ij}(\alpha )$ as the basis
functions for $\overline{\bm \Delta}$. The Wannier functions would
indeed be basis functions for $\overline{\bm \Delta}$ if they would
not be moved from one position $\bm \rho_i$ to another by some space
group operations.  Hence, we may expect that the matrices ${\bf
  D}(\alpha )$ satisfy the multiplication rule in Eq.~\refg{eq:28}
just as the matrices $\overline{\bm \Delta}(\alpha )$
do. Nevertheless, the multiplication rule should be verified
numerically in any case.
  
When using this Eq.~\refg{eq:49} a little complication arises if the
group of position $G_{0p}$ contains so few elements that the two
one-dimensional representations $\bm d_A$ and $\bm d_B$ subduced from
$\overline{\bm \Delta}$ are equal. Thus, in this case we have no
problem with the distribution of $\bm d_A$ and $\bm d_B$ across the
positions $\bm \rho_i$. Theorem~\ref{theorem2} is necessary {\em and}
sufficient and we may directly solve Eq.~\refg{eq:25} of
Theorem~\ref{theorem1}.

However, when in Sec.~\ref{sec:mgroups} or in
Sec.~\ref{sec:timeinversion} we will consider magnetic groups, we need
all the representatives ${\bf D}(\alpha )$ of the representation $\bm
D$ explicitly. Fortunately, also when the representations $\bm
  d_A$ and $\bm d_B$ are equal, Eq.~\refg{eq:49} is applicable:
in this case their exists at least one diagonal matrix representative
$\overline{\bm \Delta}(\gamma )$ of $\overline{\bm \Delta}$ with
vanishing trace and $\gamma \notin G_{0p}$.  We may define pairs
\begin{equation}
  \label{eq:50}
  (\bm \rho_a, \bm \rho_b), \quad(\bm \rho_c, \bm \rho_d),\quad \ldots
\end{equation}
of positions $\rho_i$ where the positions in each pair are connected
by the space group operation $\{\gamma | \bm t_{\gamma}\}$. In the
simplest case, we receive two pairs. Then in Eq.~\refg{eq:49} we may
identify the two representations at $\bm \rho_a$ and $\bm \rho_b$ by
$\bm d_A$ and the representations at the other two positions $\bm
\rho_c$ and $\bm \rho_d$ by $\bm d_B$. If we find four pairs of
positions, we may look for a second matrix representative
$\overline{\bm \Delta}(\gamma' )$ in $\overline{\bm \Delta}$ with
vanishing trace and $\gamma' \notin G_{0p}$. Then we may repeat the
above procedure and receive again four pairs of positions. Now we
associate the two representations $\bm d_A$ and $\bm d_B$ to the
positions $\bm \rho_i$ under the provision that always positions of
the same pair are associated with the same representation $\bm d_A$ or
$\bm d_B$.

Finally, it should be mentioned that the elements of the non-diagonal
matrices ${\bf D}(\alpha )$ are not fully fixed (as already remarked
in Ref.~\cite{ew2}): In Eq.~\refg{eq:42} we may use the elements
$\delta (\alpha )$ of {\em any} one-dimensional representation $\bm
\delta$ subducing the representation $\bm d$. We receive in each case
the same diagonal, but different non-diagonal matrices nevertheless
satisfying the multiplication rule~\refg{eq:28}. Analogously, in
Eq.~\refg{eq:49} we may determine the matrices ${\bf D}(\alpha )$ by
means of any two-dimensional representation $\overline{\bm \Delta}$
subducing $\bm d_A$ and $\bm d_B$.

In the following Theorem~\ref{theorem3} we summarize our results in
the present Sec.~\ref{sec:sufficient}.
\begin{thm}
\label{theorem3}
The Wannier function $w_i(\bm r - \bm R - \bm\rho_i)$ at the position
$\bm \rho_i$ is basis function for a one-dimensional representation
$\bm d_i$ of the ``point group of position'' $G_{0p}\subset H_0$
[Definition~\ref{def:11}], cf. Eq.~\refg{eq:40}.  The representations
$\bm d_i$ fix the (generally reducible) representation $\bm D$ of
$H_0$ defining the Wannier functions [Definition~\ref{def:4}]. The
matrix representatives ${\bf D}(\alpha )$ of $\bm D$ are unitary
generalized permutation matrices.  We distinguish between two cases
(i) and (ii).

Case (i): If the representations $\bm d_i$ are subduced from
one-dimensional representations of the point group $H_0$, then all the
Wannier functions of the band under observation are basis functions
for the same representation $\bm d$ which may be any one-dimensional
representation of $G_{0p}$ subduced from a one-dimensional
representation of $H_0$. The representation $\bm D$ exists always, its
matrix representatives ${\bf D}(\alpha )$ may be calculated by
Eq.~\refg{eq:42}.

Case (ii): If the representations $\bm d_i$ are subduced from
two-dimensional representations of the point group $H_0$, then the
Wannier functions are basis functions for the two one-dimensional
representations $\bm d_A$ and $\bm d_B$ of $G_{0p}$ subduced from the
same two-dimensional representation of $H_0$.  One half of the Wannier
functions is basis function for $\bm d_A$ and the other half for $\bm
d_B$.  In special cases, the representations $\bm d_A$ and $\bm d_B$
may be equal, see above. The representation $\bm D$ exists for a given
distribution of the representations $\bm d_A$ and $\bm d_B$ across the
positions $\bm \rho_i$ if Eq.~\refg{eq:49} yields unitary generalized
permutation matrices ${\bf D}(\alpha )$ satisfying the multiplication
rule in Eq.~\refg{eq:28}.

A third case with representations $\bm d_i$ subduced from
one-dimensional as well as from two-dimensional representations of
$H_0$ does not exist.
\end{thm}

\subsection{Not all the atoms are connected by symmetry}
\label{sec:atomsnotconnected}
If not all the atoms at the positions $\bm \rho_i$ are connected by
symmetry [Definition~\ref{def:8}], the representation $\bm D$ defining
the Wannier functions consists of representatives $\text{\bf D}(\alpha
)$ which may be written in block-diagonal form
\begin{equation}
  \label{eq:15}
{\bf D}(\alpha ) =
\left(
\begin{array}{ccc}  
\left(
\begin{array}{c}
\\
\textnormal{block 1}\\
\\
\end{array}
\right)
& 0 & \cdots\\
0 &
\left(
\begin{array}{c}
\\
\textnormal{block 2}\\
\\
\end{array} 
\right)
& \cdots\\
\vdots & \vdots \\
\end{array}
\right),
\end{equation}
where each block comprises the matrix elements $D_{ij}(\alpha )$
belonging to positions connected by symmetry. Otherwise, when the
matrices ${\bf D}(\alpha )$ would not possess block-diagonal form,
Eq.~\refg{eq:6} would falsely connect atomic positions that are not at
all connected by symmetry. As a consequence of the block-diagonal
form, the representation $\bm D$ is the direct sum over
representations $\bm D^q$ related to the individual blocks,
\begin{eqnarray}
  \label{eq:117}
  \bm D & = & \bm D^1 \oplus \bm D^2 \oplus \ldots\nonumber\\ 
        & = & \sum_q\bm D^q.
\end{eqnarray}

We may summarize as follows. 
\begin{thm}
\label{theorem8}
Each block $\bm D^q$ in Eq.~\refg{eq:117} forms its own  representation of
$H_0$ and, hence, must comply separately and independently with the
criteria given in Theorem~\ref{theorem3}.
\end{thm}

The groups of position $G_{p}$ belonging to different blocks may (but
need not) be different.  However, we assume that the sum in
Eq.~\refg{eq:117} consists only of blocks with coinciding groups of
position. If this is not true in special cases, the number $\mu$ of
the atoms in Eq.~\refg{eq:2} must be reduced until the groups of
position coincide in the sum in Eq.~\refg{eq:117}. Briefly speaking,
in such a (probably rare) case atoms of the same sort must be treated
like different atoms.

\setcounter{equation}{0}
\setcounter{definition}{0}
\setcounter{thm}{0}

\section{Condition for best localized symmetry-adapted Wannier functions}
\label{sec:condition}
Remember that we consider a closed energy band of $\mu$ branches and
let be given a representation $\bm D$ defining the Wannier functions
which was determined according to Theorems~\ref{theorem3}
and~\ref{theorem8}.  Then we may give a simple condition for best
localized symmetry-adapted Wannier functions based on the theory of
Wannier functions published in Refs.~\cite{ew1} and~\cite{ew2}.

\begin{thm}
\label{theorem1}
Let be $\bm k$ a point of symmetry in the first domain of the
Brillouin zone for the considered material and let be $H_{\bm k}
\subset H$ the little group of $\bm k$ in Herrings sense. That means,
$H_{\bm k}$ is the FINITE group denoted in Ref.~\cite{bc} by
$^H\!G^{\bm k}$ (and listed for all the space groups in Table 5.7
ibidem). Furthermore, let be $\bm D_{\bm k}$ the $\mu$-dimensional
representation of $H_{\bm k}$ whose basis functions are the $\mu$
Bloch functions $\varphi_{\bm k,q}(\bm r)$ with wave vector $\bm k$,
and $\chi_{\bm k} (a)$ (with $a \in H_{\bm k}$) the character of $\bm
D_{\bm k}$. $\bm D_{\bm k}$ either is irreducible or the direct sum
over small irreducible representations of $H_{\bm k}$.

We may choose the coefficients $g_{iq}(\bm k)$ in Eq.~\refg{eq:2} in
such a way that the Wannier functions are best localized
[Definition~\ref{def:2}] and symmetry-adapted to $H$
[Definition~\ref{def:15}] if the character $\chi_{\bm k} (a)$ of $\bm
D_{\bm k}$ satisfies at each point $\bm k$ of symmetry in the first
domain of the Brillouin zone the equation
\begin{equation}
  \label{eq:25}
\chi_{\bm k} (a) =  e^{-i\alpha\bm k\cdot\bm t_{\alpha}}~
\sum_{i = 1}^{\mu}n_{i}(a)e^{-i\bm \rho_{i}\cdot (\bm k - \alpha\bm
  k)} \text{ for } a \in H_{\bm k},
\end{equation}
where
$a = \{\alpha |\bm t_{\alpha}\}$ and
\begin{equation}
  \label{eq:32}
n_{i}(a) = \left\{
\begin{array}{ll}
d_i(\alpha ) & \text{ if }\quad \alpha \in G_{0p}\\
0 & \text{ else.}\\
\end{array}
\right.
\end{equation}
The complex numbers $d_i(\alpha )$ stand for the elements of the
one-dimensional representations $\bm d_{i}$ of $G_{0p}$ fixing the
given $\mu$-dimensional representation $\bm D$ defining the Wanner
functions.
\end{thm}

\begin{definition}[point of symmetry]
\label{def:19}
The term ``point of symmetry'' we use as defined in Ref.~\cite{bc}:
$\bm k$ is a point of symmetry if there exists a neighborhood of $\bm
k$ in which no point except $\bm k$ has the symmetry group $H_{\bm k}$. 

Thus, a point $\bm k$ of symmetry has a higher symmetry than all
surrounding points. 
\end{definition}

We add a few comments on Theorem~\ref{theorem1}.
\begin{itemize}
\item In Eq.~\refg{eq:32} we write $n_i(a)$ rather than $n_i(\alpha)$
  because the group $G_{0p}$ depends on $a = \{\alpha|\bm
  t_{\alpha}\}$. 
\item The representation $\bm D$ defining the Wannier functions is
  equivalent to the representation $\bm D_{\bm 0}$, i.e., to the
  representation $\bm D_{\bm k}$ for $\bm k = \bm 0$, see
  Eq.~\refg{eq:29}.
\item In the majority of cases all the representations $\bm d_i$
  in Eq.~\refg{eq:32} are equal. The only exceptions arises when
\begin{enumerate}
\item not all the positions $\bm \rho_i$ are connected by symmetry
  or
\item the one-dimensional representations $\bm d_i$ of $G_{0p}$
  are subduced from a higher-dimensional representation of $H_0$.
\end{enumerate}
\item A basic form of Theorem~\ref{theorem1} was published first in
  Eq.~(23) of Ref.~\cite{josla2cuo4} and used in several former
  papers. Eq.~(23) of Ref.~\cite{josla2cuo4} yields the same results as
  Theorem~\ref{theorem1}
\begin{enumerate}
\item if all the $\bm \rho_i$ are connected by symmetry and
\item if all the representations $\bm d_i$ of $G_{0p}$ are
  subduced from one-dimensional representations of $H_0$.
\end{enumerate}
These two conditions were satisfied in our former papers. 
\item The irreducible representations of the Bloch functions of the
  considered band at the points $\bm k$ of symmetry may be determined
  from the representations $\bm D_{\bm k}$ as follows:
\end{itemize}
\begin{thm}
\label{punktk}
Let $H_{\bm k}$ possess $r$ irreducible representations with the
characters $\chi_{\bm k, m}(a)\quad (1 \leq m \leq r)$ and assume that
$\bm D_{\bm k}$ contains the $m$th irreducible representation, say,
$c_{m}$ times.  Then the numbers $c_{m}$ may be calculated by means of
Eq.~(1.3.18) of Ref.~\cite{bc},
\begin{equation}
  \label{eq:88}
  c_{m} = \frac{1}{|H_{\bm k}|}\sum^{H_{\bm k}}_{a}\chi_{\bm k, m} (a)\chi^*_{\bm k} (a),
\end{equation}
where $\chi_{\bm k}(a)$ denotes the character of $\bm D_{\bm k}$ as
determined by Eq.~\refg{eq:25} and the sum runs over the $|H_{\bm k}|$
elements $a$ of $H_{\bm k}$. Remember [Theorem~\ref{theorem1}] that
$H_{\bm k}$ is a finite group.
\end{thm}

\setcounter{equation}{0}
\setcounter{definition}{0}
\setcounter{thm}{0}

\section{Proof of Theorem~\ref{theorem1}}
\label{sec:theorem1}
The existence of best localized symmetry-adapted Wannier functions is
defined in Satz 4 of Ref.~\cite{ew1}: such Wannier functions exist in
a given closed energy band of $\mu$ branches if Eqs.~(4.28) and~(4.17)
of Ref.~\cite{ew1} are satisfied. We show in this section that the
fundamental Theorem~\ref{theorem1} complies with these two equations
if the Wannier functions meet the assumptions (i) -- (iii) in
Sec.~\ref{sec:sawf}.

\subsection{Equation (4.28) of Ref.~\cite{ew1}}
\label{sec:eq4.28}
As a first step consider Eq.~(4.28) of Ref.~\cite{ew1} stating that
best localized and symmetry-adapted Wannier functions may exist only
if two representations $\widehat{\bm D}_{\bm k'_{\Sigma R}}$ and $\bm
D_{\bm k'_{\Sigma R}}$ are equivalent, 
\begin{equation}
  \label{eq:41}
  \widehat{\bm D}_{\bm k} \text{ equivalent to }
  \bm D_{\bm k},
\end{equation}
where we have abbreviated $\bm k'_{\Sigma R}$ by $\bm k$ denoting a
point of symmetry lying in the first domain of the Brillouin zone.
Consequently, our first task will be to determine the character of
$\widehat{\bm D}_{\bm k}$ as well as of $\bm D_{\bm k}$,

The representation $\bm D_{\bm k}$ as defined in
Theorem~\ref{theorem1} is the direct sum of the representations of the
Bloch functions of the considered band at point $\bm k$. The character
$\chi_{\bm k} (a)$ of the representation $\bm D_{\bm k}$ is simply
given by
\begin{equation}
\label{eq:4}
\chi_{\bm k} (a) = \text{trace\ \bf D}_{\bm k}(a)  
\end{equation}
where the matrices $\text{\bf D}_{\bm k}(a)$ are the matrix representatives
of $\bm D_{\bm k}$. 

The matrix representatives $\widehat{\text{\bf D}}_{\bm
  k}(a)$ of $\widehat{\bm D}_{\bm k}$ are defined in Eq.~(4.26) of
Ref.~\cite{ew1},
\begin{equation}
  \label{eq:31}
  \widehat{\text{\bf D}}_{\bm
    k}(a) = \text{\bf S}^*(\bm K_{\alpha})
   \text{\bf D}_{\bm 0}(\alpha )e^{-i\alpha \bm k \bm t_{\alpha}}
\end{equation}
where
\begin{equation}
  \label{eq:33}
  \bm K_{\alpha} = \bm k - \alpha \bm k
\end{equation}
is a vector of the reciprocal lattice. Again we have abbreviated $\bm
k'_{SR}$ by $\bm k$ denoting a point of symmetry. The matrices
$\text{\bf S}(\bm K)$ as defined in Eq.~(4.13) of Ref.~\cite{ew1} are
responsible for a continuous transition of the generalized Bloch
functions between neighboring Brillouin zones. The matrices ${\bf
  D}_{\bm 0}(\alpha )$ are the matrix representatives of the
representation $\bm D_{\bm k}$ for $\bm k = \bm 0$ as defined in
Theorem~\ref{theorem1}. $\bm D_{\bm 0}$ is the direct sum of the
irreducible representations of the Bloch functions of the considered
band at point $\Gamma$.

The traces of the matrices $\widehat{\text{\bf D}}_{\bm k}(a)$ can be
determined by transforming Eq.~\refg{eq:31} with the complex conjugate
of the matrix {\bf M} defined by Eq. (2.1) of Ref.~\cite{ew2},
\begin{eqnarray}
\label{eq:24}  
{\bf M}^*\widehat{\text{\bf D}}_{\bm k}(a){\bf M}^{*-1} & = &
{\bf M}^*{\bf S}^*(\bm K_\alpha){\bf M}^{*-1}\times\nonumber\\
&&{\bf M}^*{\bf D}_{\bm 0}(\alpha){\bf M}^{*-1}\times\\
&&e^{-i\alpha\bm{k\cdot t_{\alpha}}},\nonumber
\end{eqnarray}
where $a = \{\alpha|\bm t_{\alpha}\}$ still denotes an element of the space
group $H$. By definition, the matrix {\bf M} diagonalizes the matrices
$\text{\bf S}(\bm K)$ which is possible since all the $\text{\bf
  S}(\bm K)$ commute. Thus, the first factor $\text{\bf M}^*\text{\bf
  S}^*(\bm K_\alpha)\text{\bf M}^{*-1}$ in Eq.~\refg{eq:24} is the
diagonal matrix
\begin{equation}
  \label{eq:23}
  \overline{\text{\bf S}}^*(\bm K_\alpha) = 
  e^{-i\bm K_\alpha\cdot\overline{\text{\bf T}}},
\end{equation}
where, according to Eq.~(2.7) of Ref.~\cite{ew2}, also
$\overline{\text{\bf T}}$ is a diagonal matrix with
\begin{equation}
  \label{eq:27}
  \overline{\bm T}_{ii} = \bm \rho_i.
\end{equation}
Hence, the first factor in Eq.~\refg{eq:24} may be written as 
\begin{equation}
  \label{eq:38}
\begin{array}{l}
  \text{\bf M}^*\text{\bf S}^*(\bm K_\alpha)\text{\bf
    M}^{*-1} = \overline{\text{\bf S}}^*(\bm K_\alpha) =\\
\\ 
 \left(
  \begin{array}{cccc} 
  e^{-i\bm\rho_{\mu}\cdot (\bm k - \alpha\bm k)} & \ldots & 0 & 0\\              
  0 & \ddots & 0 & 0\\              
  0 & \ldots & e^{-i\bm\rho_{2}\cdot (\bm k - \alpha\bm k)} & 0 \\              
  0 & \ldots & 0 & e^{-i\bm\rho_{1}\cdot (\bm k - \alpha\bm k)}\\              
  \end{array}
  \right).\\\\
\end{array}
\end{equation}
\begin{definition}[horizontal bar]
\label{def:12}
In line with Ref.~\cite{ew2}, we denote matrices transformed with
${\bf M}$ (or ${\bf M}^*$) by a horizontal bar to indicate that these
matrices belong to the {\em diagonal} matrices $\overline{\text{\bf
    S}}(\bm K)$.
\end{definition}

As shown in Ref.~\cite{ew2} (see Eqs.~(2.18) and (2.19) of
Ref.~\cite{ew2}), the second factor
\begin{equation}
  \label{eq:141}
   \overline{\bf D}_{\bm 0}(\alpha ) = {\bf M}^*{\bf
    D}_{\bm 0}(\alpha){\bf M}^{*-1} 
\end{equation}
in Eq.~\refg{eq:24} is a matrix representative ${\bf D}(\alpha )$
of the representation $\bm D$ defining the Wannier functions,
\begin{equation}
  \label{eq:29}
  \overline{\bf D}_{\bm 0}(\alpha ) = \text{\bf D}(\alpha ).
\end{equation}

Thus, the matrices
\begin{eqnarray}
  \label{eq:126}
\overline{\widehat{{\bf D}}}_{\bm k}(a) & = & {\bf M}^*\widehat{\text{\bf D}}_{\bm k}(a){\bf M}^{*-1}\nonumber\\
& = & \overline{\bf S}^*(\bm K_\alpha){\bf D}(\alpha)
e^{-i\alpha\bm{k\cdot t_{\alpha}}}
\end{eqnarray}
are the matrix representatives of a representation
$\overline{\widehat{\bm D}}_{\bm k}$ equivalent to $\widehat{\bm
  D}_{\bm k}$.

The character of $\overline{\widehat{\bm D}}_{\bm k}$ may be easily
determined: The diagonal elements $d_i(\alpha )$ of the matrices
$\text{\bf D}(\alpha )$ are fixed by
Theorems~\ref{theorem3} and~\ref{theorem8}.  Since the
matrix $\overline{\text{\bf S}}^*(\bm K_\alpha)$ is diagonal, the
diagonal elements $\widehat{d}_i(a)$ of the matrices
$\overline{\widehat{{\bf D}}}_{\bm k}(a)$ may be written as
\begin{equation}
  \label{eq:39}
  \widehat{d}_i(a) = e^{-i\alpha\bm k\cdot\bm t_{\alpha}}
  d_i(\alpha )e^{-i\bm \rho_{i}\cdot (\bm k - \alpha\bm k)}\mbox{ for }a 
   \in H_{\bm k},
\end{equation}
where still $a = \{\alpha|\bm t_{\alpha}\}$. The diagonal elements
$d_i(\alpha )$ of the matrices $\text{\bf D}(\alpha )$ vanish if
$\alpha \notin G_{0p}$, see Eq.~\refg{eq:12}. Hence, the term on the
right-hand side of Eq.~\refg{eq:25} is the sum over the diagonal
elements $\widehat{d}_i(a)$, i.e., it is the trace of the matrices
$\overline{\widehat{{\bf D}}}_{\bm k}(a)$. Consequently, if
Eq.~\refg{eq:25} is satisfied then condition~\refg{eq:41} is true.

Strictly speaking, in Ref.~\cite{ew1} we have proven that the
condition~\refg{eq:41} must be satisfied for the points of
symmetry lying in the first domain on the {\em surface} of the
Brillouin zone.  Eq.~\refg{eq:25} demands that in addition the
representation $\bm D_{\bm 0}$ is equivalent to the representation $\bm D$
which is evidently true, see Eq.~\refg{eq:29}.

\subsection{Equation (4.17) of Ref.~\cite{ew1}}
\label{sec:eq4.17}
As a second step we show that Eq.~(4.17) of Ref.~\cite{ew1} does not
reduce the validity of Theorem~\ref{theorem1} but this equation is
satisfied whenever the assumptions (i) -- (iii) in Sec.~\ref{sec:sawf}
are valid. Taking the complex conjugate of Eq.~(4.17) of
Ref.~\cite{ew1} and transforming this equation with the matrix
$\text{\bf M}^*$ already used in Eq.~\refg{eq:24}, we receive the
equation
\begin{equation}
  \label{eq:34}
  \overline{\bf S}^*(\alpha\bm K) = {\bf D}(\alpha ) 
\overline{\bf S}^*(\bm K) {\bf D}^{-1}(\alpha )
e^{-i\alpha\bm K\cdot\bm t_{\alpha}},
\end{equation}
cf. Eqs.~\refg{eq:38} and~\refg{eq:29}, which must be satisfied for all $a
= \{\alpha |\bm t_{\alpha}\} \in H$ and all the vectors $\bm K$ of the
reciprocal lattice.

Just as the matrix
\begin{equation}
  \label{eq:37}
\overline{\text{\bf S}}^*(\bm K) =  
 \left(
  \begin{array}{cccc} 
  e^{-i\bm\rho_{\mu}\cdot \bm K} & \ldots & 0 & 0\\              
  0 & \ddots & 0 & 0\\              
  0 & \ldots & e^{-i\bm\rho_{2}\cdot \bm K} & 0\\              
  0 & \ldots & 0 & e^{-i\bm\rho_{1}\cdot \bm K}\\              
  \end{array}
  \right),
\end{equation}\\
also the matrix
$\text{\bf D}(\alpha ) \overline{\text{\bf S}}^*(\bm K) \text{\bf
  D}^{-1}(\alpha )$ in Eq.~\refg{eq:34} is diagonal with the same
diagonal elements which, however, may stand in a new order. In fact, if $D_{ji}(\alpha )
\neq 0$, the element $e^{-i\bm\rho_{i}\cdot \bm K}$ of 
$\overline{\text{\bf S}}^*(\bm K)$
at position $i$ stands at position
$j$ in the matrix $\text{\bf D}(\alpha ) \overline{\text{\bf S}}^*(\bm
K) \text{\bf D}^{-1}(\alpha )$. Thus, from Eq.~\refg{eq:34} we receive
the $\mu$ equations
\begin{equation}
  \label{eq:35}
  e^{-i\alpha\bm K\cdot\bm \rho_j}  = 
  e^{-i\bm K\cdot \bm \rho_i}\cdot e^{-i\alpha\bm K\cdot\bm t_{\alpha}}
  \ \text{ if }\ D_{ji}(\alpha ) \neq 0,
\end{equation}
yielding $\mu$ equations for the positions $\bm \rho_i$,
\begin{equation}
  \label{eq:36}
  \bm \rho_j = \alpha\bm \rho_i + \bm t_{\alpha} + \bm R_j\ 
   \text{ if }\ D_{ji}(\alpha ) \neq 0,
\end{equation}
where $\bm R_j$ is a lattice vector which may be different in each
equation. In fact, these last $\mu $ equations~\refg{eq:36} are
satisfied, see Eq.~\refg{eq:43}.

\setcounter{equation}{0}
\setcounter{definition}{0}
\setcounter{thm}{0}

\section{Magnetic groups}
\label{sec:mgroups}
Assume a magnetic structure to be given in the considered material and
let be
\begin{equation}
  \label{eq:17}
  M = H + K\{\gamma | \bm \tau \}H
\end{equation}
the magnetic group of this magnetic structure, where 
\begin{equation}
  \label{eq:98}
  \{\gamma | \bm \tau \} \in G
\end{equation}
and $K$ denotes the operator of time inversion acting on a function
$f(\bm r )$ of position according to
\begin{equation}
  \label{eq:70}
Kf(\bm r) = f^*(\bm r).  
\end{equation}

We demand that the equation
\begin{equation}
  \label{eq:30}
Kw_i\big(\gamma^{-1}(\bm r - \bm R - \bm\rho_i)\big) = \sum_{j =
  1}^{\mu}N_{ji}w_j(\bm r - \bm R - \bm\rho_i) 
\end{equation}
is satisfied in addition to Eq.~\refg{eq:6}, where the matrix
\text{$\text{\bf N} = [N_{ij}]$} is the representative of the
anti-unitary symmetry operation $K\gamma$ in the co-representation of
the point group
\begin{equation}
  \label{eq:54}
  M_0 = H_0 + K\gamma H_0  
\end{equation}
of $M$ derived from~\cite{bc} the representation $\bm D$ of $H_0$
defining the Wannier functions.

Still we assume that there is exactly one Wannier function at
each position $\bm \rho_i$, i.e., the three assumptions (i) -- (iii)
of Sec.~\ref{sec:sawf} remain valid. Thus~\cite{enhm}, Eq.~\refg{eq:30}
may be written in the more compact form
\begin{equation}
  \label{eq:57}
  KP(\{\gamma | \bm \tau \})w_{\bm T}(\bm r) = N_{ji} w_{\bm T'}(\bm r)
\end{equation}
with
\begin{equation}
  \label{eq:58}
  \bm T' = \gamma\bm T + \bm\tau
\end{equation}
and the subscripts $i$ and $j$ denote the number of the atoms at
position $\bm T$ and $\bm T'$, respectively, see
Definition~\ref{def:7}.  

\begin{definition}[symmetry-adapted to a magnetic group]
\label{def:13}
We call the Wannier functions symmetry-adapted to the magnetic group
$M$ if, in addition to Eq.~\refg{eq:8}, Eq.~\refg{eq:57} is satisfied.
\end{definition}

Again (cf. Sec.~\ref{sec:sawf}),
Eq.~\refg{eq:57} defines the non-vanishing elements of the Matrix {\bf
  N}. Hence, also {\bf N} has one non-vanishing element in each row
and each column,
\begin{equation}
  \label{eq:59}
  |N_{ji}| = 
\left\{
\begin{array}{ll}
1 & \text{if } \gamma\bm \rho_i + \bm \tau = \bm \rho_j  + \bm R \\
0 & \text{else.}\\
\end{array}
\right.
\end{equation}

As already expressed by Eq.~\refg{eq:30}, we only consider bands of
$\mu$ branches which are not connected to other bands also after the
introduction of the new anti-unitary operation $K\{\gamma | \bm \tau
\}$. That means that the considered band consists of $\mu$ branches as
well after as before the introduction of $K\{\gamma | \bm \tau
\}$. Hence, the matrix ${\bf N}$ must satisfy the equations
\begin{equation}
  \label{eq:60}
  {\bf NN}^* = \text{\bf D}(\gamma^2 ) 
\end{equation}
and
\begin{equation}
  \label{eq:61}
  {\bf D}(\alpha ) = {\bf ND}^*(\gamma^{-1}\alpha\gamma ){\bf N}^{-1} \text{ for } \alpha
  \in H_0, 
\end{equation}
see Eq.~(7.3.45) of Ref.~\cite{bc}. Still the matrices ${\bf D}(\alpha
)$ are the representatives of the representation $\bm D$ of $H_0$
defining the Wannier functions. In Ref.~\cite{bc} Eq.~(7.3.45) was
established for irreducible representations. However, this prove in
Sec.~7.3 {\em ibidem} shows that Eq.~(7.3.45) holds for reducible
representations, too, if Eq.~\refg{eq:60} is satisfied.

Assume Theorem~\ref{theorem1} to be satisfied in the considered energy
band and remember that then the coefficients $g_{iq}(\bm k)$ in
Eq.~\refg{eq:2} can be chosen in such a way that the Wannier functions
of this band are best localized and symmetry-adapted to $H$. In
Ref.~\cite{ew3} we have shown that the Wannier functions may even be
chosen symmetry-adapted to the magnetic group $M$ if Eq.~(7.1) of
Ref.~\cite{ew3},
\begin{equation}
  \label{eq:64}
  {\bf S}(-\gamma\bm K) = {\bf D}^*_{\bm 0}(K\gamma){\bf S}^*({\bm K}){\bf 
         D}^{*-1}_{\bm 0}(K\gamma)e^{-i\gamma \bm K\cdot \bm \tau},
\end{equation}
is valid for each vector $\bm K$ of the reciprocal lattice (which
should not be confused with the operator $K$ of time inversion). The
matrix $\text{\bf S}(\bm K)$ is defined in Eq.~(4.13) of
Ref.~\cite{ew1} and the matrix $\text{\bf D}_{\bm 0}(K\gamma )$ is the
representative of the symmetry operation $K\gamma$ in the
co-representation of $M_0$ derived from the representation $\bm D_{\bm
  0}$, i.e., from the representation $\bm D_{\bm k}$ for $\bm k = \bm
0$ as introduced in Theorem~\ref{theorem1}.

Transforming Eq.~\refg{eq:64} with the matrix $\text{\bf
  M}^*$ already used in Eq.~\refg{eq:24} and using
\begin{equation}
  \label{eq:65}
\begin{array}{rlll}
  {\bf N} & = & {\bf M}^*{\bf D}_{\bm 0}(K\gamma){\bf M}^{-1} & \text{(Eq.~(11.29) 
              of Ref.~\cite{ew3})}\\
\overline{\text{\bf S}}(\bm K) & = & \text{\bf M}\text{\bf S}(\bm K)\text{\bf
    M}^{-1}& \text{= diagonal, Eq.~\refg{eq:37}}\\
\overline{\bf S}^*(\gamma\bm K) & = & \overline{\bf S}(-\gamma\bm K) 
            & \text{(see Eq.~\refg{eq:37})}\\
\end{array}
\end{equation}
we receive an equation
\begin{equation}
  \label{eq:66}
  \overline{\bf S}^*(\gamma \bm K ) = {\bf N}^*\overline{\bf S}^*(\bm K )
               {\bf N}^{*-1}e^{-i\gamma\bm K\cdot\bm \tau}  
\end{equation}
identical to Eq.~\refg{eq:34} when we replace the space group
operation $\{\alpha |\bm t_{\alpha}\}$ by $\{\gamma |\bm \tau\}$ and
${\bf D}(\alpha )$ by ${\bf N}^*$. In Sec.~\ref{sec:eq4.17} we have
shown that Eq.~\refg{eq:34} is satisfied if the matrices ${\bf
  D}(\alpha )$ follow Eq.~\refg{eq:43}. In the same way,
Eq.~\refg{eq:66} is true if the elements of ${\bf N}$ (as well as of
${\bf N}^*$) obey Eq.~\refg{eq:59}. Thus,
Eqs.~\refg{eq:59},~\refg{eq:60}, and~\refg{eq:61} are the only
additional conditions for the existence of best localized Wannier
functions which are symmetry-adapted to the magnetic group $M$.

We summarize the results of the present Sec.~\ref{sec:mgroups} in
\begin{thm}
\label{theorem4}
The coefficients $g_{iq}(\bm k )$ in Eqs.~\refg{eq:2} may be chosen in
such a way that the Wannier functions are best localized
[Definition~\ref{def:3}] and even symmetry-adapted to the magnetic
group $M$ in Eq.~\refg{eq:17} [Definition~\ref{def:13}] if, according
to Theorem~\ref{theorem1}, they may be chosen symmetry-adapted to $H$
and if, in addition, there exists a $\mu$-dimensional matrix {\bf N}
satisfying Eqs.~\refg{eq:59},~\refg{eq:60} and~\refg{eq:61}.

The representation $\bm D$ in Eqs.~\refg{eq:60} and~\refg{eq:61} is
the representation defining the Wannier functions as used in
Theorem~\ref{theorem1}.

In most cases, we may put the
non-vanishing elements of {\bf N} equal to 1.
\end{thm}
\begin{definition}[magnetic band]
\label{def:21}
If, according to Theorem~\ref{theorem4}, the unitary transformation in
Eq.~\refg{eq:1} may be chosen in such a way that the Wannier functions
are best localized and symmetry-adapted to the magnetic group $M$ in
Eq.~\refg{eq:17}, we call the band under consideration [as defined by the
representations $\bm D_{\bm k}$ in Eq.~\refg{eq:25}] a
``magnetic band related to the magnetic group $M$''.

Within the nonadiabatic Heisenberg model, the existence of a narrow,
roughly half-filled magnetic band in the band structure of a material
is a precondition for the stability of a magnetic structure with the
magnetic group $M$ in this material. However, the magnetic group $M$
must be ``allowed'' in order that the time-inversion symmetry does not
interfere with the stability of the magnetic state~\cite{bafe2as2}.
\end{definition}

\setcounter{equation}{0}
\setcounter{definition}{0}
\setcounter{thm}{0}

\section{Spin-dependent Wannier functions}
\label{sec:sdwf}

\subsection{Definition}
\label{sec:defsdwf}
Assume the Hamiltonian $\mathcal{H}$ of a single electron in the
considered material to be given and assume $\mathcal{H}$ to consist of a
spin-independent part $\mathcal{H}_i$ and a spin-dependent perturbation
$\mathcal{H}_s$,
\begin{equation}
  \label{eq:87}
  \mathcal{H} = \mathcal{H}_i + \mathcal{H}_s.
\end{equation}
Further assume the Bloch spinors $\psi_{\bm k,q,s}(\bm r, t)$ as the
exact solutions of the Schr\"odinger equation 
\begin{equation}
  \label{eq:109}
  \mathcal{H}\psi_{\bm k,q,s}(\bm r, t) = E_{\bm k,q,s}\psi_{\bm k,q,s}(\bm r, t)
\end{equation}
to be completely determined in the first domain of the Brillouin
zone. Just as the Bloch functions, they are labeled by the wave vector
$\bm k$ and the branch index $q$. In addition, they depend on the spin
coordinate $t = \pm\frac{1}{2}$ and are labeled by the spin quantum
number $s = \pm\frac{1}{2}$.

Consider again a closed energy band of $\mu$ branches which, in
general, was not closed before the perturbation $\mathcal{H}_s$ was
activated.  Now each branch is doubled, that means that it consists of
two bands related to the two different spin directions.  Just as in
Sec.~\ref{sec:defwf} we assume that the Bloch spinors $\psi_{\bm
  k,q,s}(\bm r, t)$ are chosen in such a way that they vary
continuously through the first domain and approach continuously the
boundaries of the first domain. In the rest of the Brillouin zone and
in the remaining $\bm k$ space they shall be given again by
Eqs.~\refg{eq:95} and~\refg{eq:96}~\cite{ew3} where, however, $P(a)$
acts now on both $\bm r$ and $t$, see Eq.~\refg{eq:78}.

We define ``spin-dependent Wannier functions'' by replacing the Bloch
functions $\varphi_{\bm k,q}(\bm r)$ in Eq.~\refg{eq:2} by linear
combinations
\begin{equation}
  \label{eq:62}
  \varphi_{\bm k,q,m}(\bm r,t) = \sum_{s = -\frac{1}{2}}^{+\frac{1}{2}}
f_{ms}(q,\bm k)\psi_{\bm k,q,s}(\bm r, t)
\end{equation}
of the given Bloch spinors. Thus, Eq.~\refg{eq:2} becomes
\begin{equation}
  \label{eq:67}
  \widetilde\varphi_{\bm k,i,m}(\bm r, t) = 
  \sum_{q = 1}^{\mu}g_{iq}(\bm k
  )\varphi_{\bm k,q,m}(\bm r, t)
\end{equation}
and, finally, the spin-dependent Wannier functions my be written as 
\begin{equation}
  \label{eq:68}
  w_{i,m}(\bm r - \bm R - \bm\rho_i, t) = \frac{1}{\sqrt{N}}\sum^{BZ}_{\bm
  k}e^{-i\bm k (\bm R + \bm\rho_i)}\widetilde\varphi_{\bm k,i,m}(\bm
r, t).
\end{equation}
Also the spin-dependent Wannier functions depend on $t$ and are
labeled by a new quantum number $m = \pm\frac{1}{2}$ which, in the
framework of the nonadiabatic Heisenberg model, is interpreted as the
quantum number of the ``crystal spin''~\cite{es3,es,enhm}. The sum in
Eq.~\refg{eq:67} runs over the $\mu$ branches of the given closed
energy band, where $\mu$ still is equal to the number of the
considered atoms in the unit cell.

The matrices
\begin{equation}
  \label{eq:148}
{\bf g}(\bm k ) = [g_{iq}(\bm k )]  
\end{equation}
still are unitary [see Eq.~\refg{eq:5}] and also the coefficients
$f_{ms}(q,\bm k)$ in Eq.~\refg{eq:62} form for each $\bm k$ and $q$ a
two-dimensional matrix
\begin{equation}
  \label{eq:94}
  {\bf f}(q, \bm k) = [f_{ms}(q,\bm k)]  
\end{equation}
which is unitary,
\begin{equation}
  \label{eq:86}
{\bf f}^{-1}(q, \bm k) = {\bf f}^{\dagger}(q, \bm k),  
\end{equation}
in order that the spin-dependent Wannier functions are
orthonormal,
\begin{equation}
  \label{eq:53}
\begin{array}{c}
  \displaystyle\sum^{+\frac{1}{2}}_{t = -\frac{1}{2}}\int w^*_{i,m}(\bm r - \bm R - \bm\rho_i,
  t)w_{i',m'}(\bm r - \bm R' 
  - \bm\rho_{i'}, t)d\bm r\\\\
  = \delta_{\bm R\bm R'}\delta_{ii'}\delta_{mm'}.
\end{array}
\end{equation} 

Within the nonadiabatic Heisenberg model we strictly consider the
limiting case of vanishing spin-orbit coupling,
\begin{equation}
  \label{eq:55}
  \mathcal{H}_s \rightarrow 0,
\end{equation}
by approximating the Bloch spinors $\psi_{\bm k,q,s}(\bm r, t)$ by
means of the spin-independent Bloch functions $\varphi_{\bm k,q}(\bm
r) $. In this context, we should distinguish between two kinds of
Bloch states $\varphi_{\bm k,q}(\bm r)$ in the considered closed band:
\begin{enumerate}
\item If $\varphi_{\bm k,q}(\bm r)$ 
  \begin{itemize}
  \item was basis function for a non-degenerate representation already
    before the spin-dependent perturbation $\mathcal{H}_s$ was
    activated, or
\item was basis function for a degenerate representation before
  $\mathcal{H}_s$ was activated, and this degeneracy is not removed by
  $\mathcal{H}_s$ [see Sec.~\ref{sec:fkqd}],
  \end{itemize}
  then we may approximate the Bloch spinors by
  \begin{equation}
    \label{eq:56}
    \psi_{\bm k,q,s}(\bm r, t) = u_s(t)\varphi_{\bm
      k,q}(\bm r) 
  \end{equation}
where the functions $u_{s}(t)$ denote Pauli's spin
functions
\begin{equation}
  \label{eq:63}
  u_{s}(t) = \delta_{st},
\end{equation}
with the spin quantum number $s = \pm\frac{1}{2}$ and the spin
coordinate $t = \pm\frac{1}{2}$. Eq.~\refg{eq:56} applies to the {\em
  vast majority} of points $\bm k$ in the Brillouin zone.
\item If at a special point $\bm k$ the Bloch function $\varphi_{\bm
    k,q}(\bm r)$ was basis function for a degenerate single-valued
  representation before the perturbation $\mathcal{H}_s$ was activated
  and if this degeneracy is removed by $\mathcal{H}_s$, then
  Eq.~\refg{eq:56} is unusable for the sole reason that we do not know
  which of the basis functions of the degenerate representation we
  should avail in this equation. In fact, in this case the Bloch
  spinors $\psi_{\bm k,q,s}(\bm r, t)$ are well defined linear
  combinations of the functions $u_s(t)\varphi_{\bm k,q}(\bm r)$
  comprising {\em all} the basis functions $\varphi_{\bm k,q}(\bm r)$
  of the degenerate single-valued representation (as given, e.g., in
  Table 6.12 of Ref.~\cite{bc}). These specific linear combinations
  are not considered because, at this stage, they are of no importance
  within the nonadiabatic Heisenberg model.
\end{enumerate}

In the framework of the approximation defined by Eq.~\refg{eq:56} the
two functions $\varphi_{\bm k,q,m}(\bm r,t)$ in Eq.~\refg{eq:62} (with
$m = \pm\frac{1}{2}$) are usual Bloch functions with the spins lying
in $\pm z$ direction if
\begin{equation}
  \label{eq:71}
f_{ms}(q,\bm k) = \delta_{ms}.  
\end{equation}
Otherwise, if the coefficients $f_{ms}(q,\bm k)$ cannot be chosen
independent of $\bm k$, the spin-dependent Wannier functions cannot be
written as a product of a local function with the spin function
$u_s(t)$ even if the approximation defined by Eq.~\refg{eq:56} is
valid. Consequently, even in the limit of vanishing spin-orbit
coupling, the spin-dependent Wannier functions are neither orthonormal
in the local space $\mathcal{L}$ nor in the spin space $\mathcal{S}$,
but in $\mathcal{L}\times\mathcal{S}$ only, see Eq.~\refg{eq:53}.
Thus, also in the case $$\mathcal{H}_s \rightarrow 0$$ spin-dependent
Wannier functions clearly differ from usual Wannier functions
characterized by $$\mathcal{H}_s = 0.$$ The ansatz~\refg{eq:68}
presents the most general definition of Wannier functions. While their
localization can be understood only in terms of the exact solutions of
the Schr\"odinger equation~\refg{eq:109}, the limiting case of
vanishing spin-orbit coupling characterized by Eq.~\refg{eq:56} yields
fundamental properties of these Wannier functions leading finally to
an understanding of the material properties of
superconductors~\cite{es1,es,josn}.

\subsection{Symmetry-adapted spin-dependent Wannier functions}
\label{sec:sasdwf}
We demand that symmetry-adapted spin-dependent Wannier functions
satisfy in analogy to Eq.~\refg{eq:8} the equation
\begin{equation}
  \label{eq:72}
 P(a)w_{\bm T, m}(\bm r, t) = D_{ji}(\alpha )\sum_{m' =
  -\frac{1}{2}}^{\frac{1}{2}} d_{m'm}(\alpha )w_{\bm T',m'}(\bm r, t) 
\end{equation}
for $a \in H$
since still the assumptions (i) -- (iii) of Sec.~\ref{sec:sawf} are
valid. Merely the third assumption (iii) is modified: now the {\em
  two} Wannier functions $w_{\bm T, +\frac{1}{2}}(\bm r, t)$ and
$w_{\bm T, -\frac{1}{2}}(\bm r, t)$ are situated at the same atom and,
consequently, we now put
\begin{equation}
  \label{eq:73}
w_{\bm T, m}(\bm r, t) \equiv w_{i,m}(\bm r - \bm R - \bm\rho_i, t),
\end{equation}
where $m = \pm\frac{1}{2}$.

The vectors $\bm T$ and $\bm T'$ are still
given by Eqs.~\refg{eq:22} and~\refg{eq:9}, respectively.  The
matrices ${\bf D}(\alpha ) = [D_{ij}(\alpha )]$ in Eq.~\refg{eq:72}
are again unitary generalized permutation matrices, and the subscripts
$i$ and $j$ denote the number of the atoms at position $\bm T$ and
$\bm T'$, respectively, see Definition~\ref{def:7}.

The operators $P(a)$ now act additionally on the spin coordinate $t$
of a function $f(\bm r, t)$,
\begin{equation}
  \label{eq:78}
  P(a)f(\bm r, t) = f(\alpha^{-1}\bm r - \alpha^{-1}\bm t_{\alpha}, \alpha^{-1}t),  
\end{equation}
where the effect of a point group operation on the
spin coordinate $t$ of the spin function $u_s(t)$ is given by the
equation~\cite{bc}
\begin{equation}
  \label{eq:79}
u_s(\alpha^{-1}t) = \sum_{s'} d_{s's}(\alpha )u_{s'}(t) \mbox{ for } \alpha \in H^d_0.
\end{equation}
The matrix
\begin{equation}
  \label{eq:80}
{\bf d}_{1/2}(\alpha ) = [d_{ss'}(\alpha )]   
\end{equation}
denotes the representative of $\alpha$ in the two-dimensional {\em
  double-valued} representation $\bm d_{1/2}$ of $O(3)$ as listed, e.g.,
in Table 6.1 of Ref.~\cite{bc}.

We have to take into consideration that the double-valued
representations of a group $g$ are not really representations of $g$
but of the abstract ``double group'' $g^d$ of order $2|g|$, while the
single valued representations are representations of both $g$ and
$g^d$~\cite{bc}.

\begin{definition}[double-valued]
\label{def:5}
Though we use the familiar expression ``double-valued''
representation of a group $g$, we consider the double-valued
representations as ordinary single-valued representations of the
related abstract double group $g^d$, denoted by a superscript ``d''.
\end{definition}

Since the index $m$ of the spin-dependent Wannier functions is
interpreted as spin quantum number, we demand that the term
$$ \sum_{m' = -\frac{1}{2}}^{\frac{1}{2}} d_{m'm}(\alpha )w_{\bm T',m'}(\bm r, t) $$
in Eq.~\refg{eq:72} describes a rotation or reflection of the crystal
spin. Thus, we demand that also the matrices $[d_{mm'}(\alpha )]$ are
the representatives of the two-dimensional double-valued
representation $\bm d_{1/2}$,
\begin{equation}
  \label{eq:77} [d_{mm'}(\alpha )] = {\bf d}_{1/2}(\alpha )\quad\text{
for } \alpha \in H^d_0.
\end{equation} 

\begin{definition}[symmetry-adapted]
\label{def:18}
We call the spin-dependent Wannier functions ``symmetry-adapted to the
double group $H^d$ related to space group $H$'' if they satisfy
Eq.~\refg{eq:72} for $a \in H^d$, where the matrices $[d_{mm'}(\alpha
)]$ are the representatives of the two-dimensional double-valued
representation $\bm d_{1/2}$ of $O(3)$.
\end{definition}

Consequently, symmetry-adapted spin-dependent Wannier
functions are basis functions for the double-valued representation
\begin{equation}
  \label{eq:75} 
  \bm D^d = \bm D \otimes \bm d_{1/2}
\end{equation} of $H^d_0$ which is the inner Kronecker product of the
single-valued representation $\bm D$ defined by Eq.~\refg{eq:72}
and the double-valued representation $\bm
d_{1/2}$. Thus, the $2\mu$-dimensional matrix representatives ${\bf
  D}^d(\alpha)$ of $\bm D^d$ may be written as Kronecker
products,
\begin{equation}
  \label{eq:76} 
{\bf D}^d(\alpha) = {\bf D}(\alpha ) \times
{\bf d}_{1/2}(\alpha ).
\end{equation}

\begin{definition}[representation defining the spin-dependent
  Wannier functions]
\label{def:17}
The single-valued representation $\bm D$ of $H_0$ defined by
Eq.~\refg{eq:72} shall be shortly referred to as ``the representation
defining the spin-dependent Wannier functions'' and its matrix
representatives $${\bf D}(\alpha ) = [D_{ij}(\alpha )]$$ to as ``the
matrices defining the spin-dependent Wannier functions''.

While usual (spin-independent) Wannier functions are basis functions
for the representation $\bm D$ defining the Wannier functions,
spin-dependent Wannier functions are basis functions for the
double-valued representation
$$\bm D^d = \bm D \otimes \bm d_{1/2}$$ in Eq.~\refg{eq:75}. 
\end{definition}

Also the representation $\bm D$ defining the spin-dependent Wannier
functions has to meet the conditions given in Sec.~\ref{sec:matrices}
as shall be summarized in

\begin{thm}
\label{theorem11}
The two spin-dependent Wannier function $w_{i,\frac{1}{2}}(\bm r - \bm
R - \bm\rho_i, t)$ and $w_{i,-\frac{1}{2}}(\bm r - \bm R - \bm\rho_i,
t)$ at the position $\bm \rho_i$ are basis functions for the
two-dimensional representation
\begin{equation}
  \label{eq:99}
  \bm d^d_i = \bm d_i \otimes \bm d_{1/2}
\end{equation}
of the double group $G^d_{0p}$ related to the point group of position
$G_{0p}$. The one-dimensional representations $\bm d_i$ in
Eq.~\refg{eq:99} fix the (generally reducible) representation $\bm D$
of $H_0$ defining the spin-dependent Wannier functions
[Definition~\ref{def:17}]. The matrix representatives ${\bf D}(\alpha
)$ of $\bm D$ still are unitary generalized permutation matrices which
must be chosen in such a way that they form a representation of $H_0$.
We again distinguish between the two cases (i) and (ii) defined in
Theorem~\ref{theorem3}.

In addition, Theorem~\ref{theorem8} must be noted.
\end{thm}

Theorem~\ref{theorem1} does not distinguish between usual and
spin-dependent Wannier functions but uses only the special
representations of the Bloch functions or Bloch spinors, respectively,
at the points $\bm k$ of symmetry.  Thus, Theorem~\ref{theorem1}
applies to both usual and spin-dependent Wannier functions if in the
case of spin-dependent Wannier functions we replace the little groups
$H_{\bm k}$ by the double groups $H^d_{\bm k}$. Just as the groups
$H_{\bm k}$, the groups $H^d_{\bm k}$ are {\em finite} groups in
Herrings sense as denoted in Ref.~\cite{bc} by $^H\!G^{\dagger\bm k}$
and, fortunately, are explicitly given in Table 6.13 {\em ibidem}.

When we consider single-valued representations, then the sum on the
right-hand side of Eq.~\refg{eq:25} runs over the $\mu$ diagonal
elements $\widehat{d}_{i}(a)$ of the matrices $\overline{\widehat{\bf
      D}}_{\bm k}(a)$ in Eq.~\refg{eq:126}. When we consider
double-valued representations, on the other hand, this sum runs over
$2\mu$ diagonal elements $\widehat{d}^d_{i,m}(a)$ of the corresponding
matrices
\begin{equation}
  \label{eq:133}
\overline{\widehat{\bf D}}^d_{\bm k}(a) = \overline{\bf S}^{d*}(\bm K_\alpha){\bf D}^d(\alpha)
e^{-i\alpha\bm{k\cdot t_{\alpha}}}
\end{equation}
where
\begin{equation}
  \label{eq:134}
  \overline{\bf S}^{d*}(\bm K_\alpha) = \overline{\bf S}^*(\bm K_\alpha) \times
\left(
\begin{array}{cc}
1 & 0\\
0 & 1
\end{array}
\right)
\end{equation}
[where $\overline{\bf S}^*(\bm K_\alpha)$ is given in Eq.~\refg{eq:38}]
because also $\overline{\bf S}^{d*}(\bm K_\alpha)$ is diagonal and
now there are two Wannier functions $w_{i,m}(\bm r - \bm R -
\bm\rho_i, t)$ with $m = \pm\frac{1}{2}$ at each position $\bm
\rho_i$.

We need not to solve Eq.~\refg{eq:25} directly but we may determine
the representations $\bm D^d_{\bm k}$ complying with Eq.~\refg{eq:25}
in a quicker way. Eq.~\refg{eq:134} shows that we may write the
matrices $\overline{\widehat{\bf D}}^d_{\bm k}(a)$ simply as Kronecker
products,
\begin{equation}
  \label{eq:142}
  \overline{\widehat{\bf D}}^d_{\bm k}(a) = \overline{\widehat{\bf
      D}}_{\bm k}(a) \times
{\bf d}_{1/2}(\alpha ),
\end{equation}
where $\overline{\widehat{\bf D}}_{\bm k}(a)$ is given in
Eq.~\refg{eq:126}.

Now assume that we have already determined according to
Theorem~\ref{theorem1} the single-valued representations $\bm
D^{\text{aff}}_{\bm k}$ in the closed band under consideration. Then,
the representations $\overline{\widehat{\bm D}}_{\bm k}$ and $\bm
D^{\text{aff}}_{\bm k} $ are equivalent [see Eq.~\refg{eq:41}] and,
consequently, also the representations
\begin{equation}
  \label{eq:143}
  \overline{\widehat{\bm
    D}}^d_{\bm k} = \overline{\widehat{\bm
      D}}_{\bm k} \otimes
{\bm d}_{1/2}
\end{equation}
and 
\begin{equation}
  \label{eq:85}
  \bm D^d_{\bm k} = \bm D^{\text{aff}}_{\bm k}  \otimes \bm d_{1/2}
\end{equation}
are equivalent. Hence [Sec.~\ref{sec:eq4.28}] the double-valued
representations $\bm D^d_{\bm k}$ comply with Theorem~\ref{theorem1}
in the same way as the single-valued representations $\bm
D^{\text{aff}}_{\bm k}$ do.

\begin{definition}[affiliated single-valued band]
\label{def:20}
In this context we call the band defined by the double-valued
representations $\bm D^d_{\bm k}$ in Eq.~\refg{eq:85} the
``double-valued band'' and the band defined by the single-valued
representations $\bm D^{\text{\em aff}}_{\bm k}$ an ``affiliated
single-valued band''.

While a double-valued band may possess several affiliated
single-valued bands, any single-valued band is affiliated to exactly
one double-valued band. 

The affiliated single-valued band is a closed band that, generally,
does not exist in the band structure of the considered material. That
means that the Bloch functions $\varphi_{\bm k,q}(\bm r)$ of the
closed band under consideration band generally do not form a basis for
the representations $\bm D^{\text{\em aff}}_{\bm k}$ even if
Eq.~\refg{eq:56} is valid, see, e.g., the single-valued band
affiliated to the superconducting band [Definition~\ref{def:22}] of
niobium as given in Eq.~\refg{eq:97}.
\end{definition}

We may summarize the result of this section in
\begin{thm}
\label{theorem5}
Remember that we consider a closed energy band of $\mu$ branches and
let be given a representation $\bm D$ defining the spin-dependent
Wannier functions which was determined according to
Theorem~\ref{theorem11}. The band may only be closed after the
spin-dependent perturbation $\mathcal{H}_s$ was activated.

Let be $\bm k$ a point of symmetry in the first domain of the
Brillouin zone for the considered material and let be $H^d_{\bm k}$
the little double group of $\bm k$ in Herrings sense. That means,
$H^d_{\bm k}$ is the FINITE group denoted in Ref.~\cite{bc} by
$^H\!G^{\dagger\bm k}$ and explicitly given in Table 6.13
ibidem. Furthermore, let be $\bm D^d_{\bm k}$ the $2\mu$-dimensional
representation of $H^d_{\bm k}$ whose basis functions are the $2\mu$
Bloch spinors $\psi_{\bm k,q,s}(\bm r, t)$ with wave vector $\bm
k$. $\bm D^d_{\bm k}$ either is irreducible or the direct sum over
double-valued irreducible representations of $H^d_{\bm k}$.  The
representations $\bm D^d_{\bm k}$ follow Eq.~\refg{eq:85},
\begin{equation}
  \label{eq:157}
  \bm D^d_{\bm k} = \bm D^{\text{\em aff}}_{\bm k}  \otimes \bm d_{1/2},
\end{equation}
where the $\mu$-dimensional representations $\bm D^{\text{\em
    aff}}_{\bm k}$ define the affiliated single-valued band. Thus,
also each $\bm D^{\text{\em aff}}_{\bm k}$ is the direct sum over
single-valued irreducible representations of $H_0$.

We may choose the coefficients $g_{iq}(\bm k )$ and $f_{ms}(q,\bm k)$
in Eqs.~\refg{eq:67} and~\refg{eq:62}, respectively, in such a way
that the spin-dependent Wannier functions are best localized
[Definition~\ref{def:3}] and symmetry-adapted to the double group
$H^d$ [Definition~\ref{def:18}] if the characters $\chi_{\bm k} (a)$
of the single-valued representations $\bm D^{\text{\em aff}}_{\bm k}$
satisfy Eq.~\refg{eq:25}.

The complex numbers $d_i(\alpha )$ in Eq.~\refg{eq:32} stand for the
elements of the one-dimensional representations $\bm d_{i}$ of
$G_{0p}$ fixing the given representation $\bm D$ defining the
spin-dependent Wannier functions [according to
Definition~\ref{def:17}].
\end{thm}

\subsection{Time inversion}
\label{sec:timeinversion}
\subsubsection{Time-inversion symmetry of the spin-dependent Wannier functions}
\label{sec:timeinversionsdwf}
Within the nonadiabatic Heisenberg model we are not interested in
spin-dependent Wannier functions that are symmetry-adapted to a
general magnetic group as given in Eq.~\refg{eq:17}, but we only
demand that they are adapted to the ``grey''~\cite{bc} magnetic group
\begin{equation}
  \label{eq:100}
  M^d = H^d + KH^d,  
\end{equation}
or, in brief, we demand that they are adapted to the time-inversion
symmetry. $K$ still denotes the operator of time inversion acting on a
function of position $f(\bm r)$ according to Eq.~\refg{eq:70} and on
Pauli's spin functions $u_s(t)$ according to
\begin{equation}
  \label{eq:101}
  Ku_s(t) = \pm u_{-s}(t)
\end{equation}
(see, e.g., Table 7.15 of Ref.~\cite{bc}), where we may define the
plus to belong to $s = +\frac{1}{2}$ and the minus to $s =
-\frac{1}{2}$.

The index $m$ of the spin-dependent Wannier functions we still
interpret as the quantum number of the crystal spin. Consequently, we
demand that $K$ acts on $m$ in the same way as it act on $s$,
\begin{equation}
  \label{eq:102}
  Kw_{\bm T, m}(\bm r, t) = \pm w_{\bm T, -m}(\bm r, t)
\end{equation}
where again we define the plus to belong to $m = +\frac{1}{2}$ and
the minus to $m = -\frac{1}{2}$.

\begin{definition}[symmetry-adapted to a magnetic group]
\label{def:6}
We call the spin-dependent Wannier functions ``symmetry-adapted to the
magnetic group $M^d$'' as given in Eq.~\refg{eq:100} if they
are symmetry-adapted to $H^d$ [Definition~\ref{def:18}], and if, in
addition, Eq.~\refg{eq:102} is satisfied.
\end{definition}

In analogy to Eq.~\refg{eq:72}, Eq.~\refg{eq:102} may be written as
\begin{equation}
  \label{eq:81}
Kw_{\bm T, m}(\bm r, t) = N_{ii}\sum_{m' =
  -\frac{1}{2}}^{\frac{1}{2}} n_{m'm}w_{\bm T,m'}(\bm r, t) 
\end{equation}
where ${\bf N} = [N_{ij}]$ denotes the $\mu$-dimensional identity
matrix
\begin{equation}
  \label{eq:105}
{\bf N} = 
   \left(
  \begin{array}{cccc} 
  1 & 0 & \ldots & 0\\              
  0 & 1 & \ldots & 0\\   
  \vdots & \vdots & \ddots & \vdots\\           
  0 & 0 & \ldots & 1\\              
  \end{array}
  \right)
= {\bf 1}
\end{equation}
and
\begin{equation}
  \label{eq:104}
  {\bf n} = [n_{mm'}] = 
 \left(
  \begin{array}{cc} 
  0 & -1 \\              
  1 & 0 \\              
  \end{array}
  \right).
\end{equation}

Eq.~\refg{eq:81} shows that the $2\mu$-dimensional
matrix
\begin{equation}
  \label{eq:103}
  {\bf N}^d = {\bf N} 
  \times
{\bf n}
\end{equation}
is the matrix representative of the operator $K$ of time inversion in
the co-representation of the magnetic point group 
\begin{equation}
  \label{eq:51}
  M^d_0 = H^d_0 + KH^d_0
\end{equation}
derived from the representation $\bm D^d$ in Eq.~\refg{eq:75}.  Thus,
the matrix ${\bf N}^d$ has to comply (Sec.~\ref{sec:mgroups}) with the
three equations ~\refg{eq:60},~\refg{eq:61} and~\refg{eq:66} which now
may be written as
\begin{align}
  \label{eq:106}
  {\bf N}^d{\bf N}^{d*} & = {\bf D}^d(K^2)  = {\bf -1},\\
  \label{eq:108}
  {\bf D}^d(\alpha ) & = {\bf N}^d{\bf D}^{d*}(\alpha ){\bf N}^{d-1} \text{ for } \alpha
  \in H^d_0,\\
  \intertext{and}
  \label{eq:107}
  \overline{\bf S}^{d*}(\bm K ) & = {\bf N}^{d*}\overline{\bf S}^{d*}(\bm K )
               {\bf N}^{d*-1},  
\end{align}
respectively.

The first Eq.~\refg{eq:106} is true because
\begin{equation}
  \label{eq:92}
  {\bf n} {\bf n}^*\ (= {\bf n} {\bf n}) = {\bf -1} 
\end{equation}
and the second Eq.~\refg{eq:108} is satisfied if 
{\bf n} and {\bf N} in Eq.~\refg{eq:103} follow two conditions,
\begin{equation}
  \label{eq:114}
    {\bf d}_{1/2}(\alpha ) = {\bf n}{\bf d}^{*}_{1/2}(\alpha
    ){\bf n}^{-1} \text{ for } \alpha
      \in H^d_0,
\end{equation}
and
\begin{equation}
  \label{eq:116}
    {\bf D}(\alpha ) = {\bf N}{\bf D}^{*}(\alpha
    ){\bf N}^{-1} \text{ for } \alpha
      \in H_0.
\end{equation}
The first condition~\refg{eq:114} is always valid, see, e.g., Table
7.15 (q) of Ref.~\cite{bc}, and the second condition~\refg{eq:116} is
satisfied if the representation $\bm D$ defining the spin-dependent
Wannier functions is real.

In the third Eq.~\refg{eq:107} the diagonal matrix $\overline{\bf
  S}^{d*}(\bm K )$ has the form
\begin{equation}
  \label{eq:10}
  \overline{\bf S}^{d*}(\bm K ) =
  \overline{\bf S}^{*}(\bm K )
\times
\left(
\begin{array}{cc}
1 & 0 \\
0 & 1
\end{array}
\right) 
\end{equation}
[cf. Eq.\refg{eq:134}] where $\overline{\bf S}^{*}(\bm K )$ is given in Eq.~\refg{eq:37}.
Thus, Eq.~\refg{eq:107} decomposes into two parts,
\begin{equation}
  \label{eq:111}
 \left(
  \begin{array}{cc} 
  1 & 0 \\              
  0 & 1 \\             
  \end{array}
  \right)
  = {\bf n}
 \left(
  \begin{array}{cc} 
  1 & 0 \\              
  0 & 1 \\             
  \end{array}
 \right)
  {\bf n}^{-1}  
\end{equation}
and
\begin{equation}
  \label{eq:112}
  \overline{\bf S}^{*}(\bm K ) = {\bf N}^{*}\overline{\bf S}^{*}(\bm K )
               {\bf N}^{*-1}  
\end{equation}
which both are evidently satisfied.

We summarize our results in this Sec.~\ref{sec:timeinversionsdwf} in
\begin{thm}
\label{theorem6}
The coefficients $g_{iq}(\bm k )$ and $f_{ms}(q,\bm k)$ in
Eqs.~\refg{eq:67} and~\refg{eq:62}, respectively, may be chosen in
such a way that the spin-dependent Wannier functions are best
localized [Definition~\ref{def:3}] and even symmetry-adapted to the
magnetic group $M^d$ in Eq.~\refg{eq:100} [Definition~\ref{def:6}] if,
according to Theorem~\ref{theorem5}, they may be chosen symmetry-adapted to $H^d$
and if, in addition, the representation $\bm D$ defining the
spin-dependent Wannier functions used in Theorem~\ref{theorem5} is
real.
\end{thm}

\begin{definition}[superconducting band]
\label{def:22}
If, according to Theorem~\ref{theorem6}, the unitary transformation in
Eq.~\refg{eq:68} may be chosen in such a way that the spin-dependent
Wannier functions are best localized and symmetry-adapted to the
magnetic group $M^d$ in Eq.~\refg{eq:100}, we call the band under
consideration [as defined by the double-valued representations $\bm
D^d_{\bm k}$ in Eq.~\refg{eq:157}] a ``superconducting band''.

Within the nonadiabatic Heisenberg model, the existence of a narrow,
roughly half-filled superconducting band in the band structure of a
material is a precondition for the stability of a superconducting
state in this material.
\end{definition}

\subsubsection{Time-inversion symmetry of the matrices ${\bf f}(q, \bm
  k)$}
\label{sec:timeinversionfqk}
In this section we derive the time-inversion symmetry of the matrices
${\bf f}(q, \bm k)$ defined in Eq.~\refg{eq:62} and shall give the
result in Theorem~\ref{theorem12}. Thought evidence for this important
theorem was already provided in Ref.~\cite{es2} and later
papers~\cite{es,josn}, we repeat the proof with the notations used in
the present paper.

Combining Eqs.~\refg{eq:67} and~\refg{eq:68} we may write the
spin-dependent Wannier functions as
\begin{equation}
  \label{eq:110}
\begin{array}{l}
  w_{i,m}(\bm r - \bm R - \bm\rho_i, t) = \\\\\displaystyle\frac{1}{\sqrt{N}}
  \sum^{BZ}_{\bm
    k}\sum_{q = 1}^{\mu}e^{-i\bm k (\bm R + \bm\rho_i)}g_{iq}(\bm k
  )\varphi_{\bm k,q,m}(\bm r, t).  
\end{array}
\end{equation}

By application of the operator $K$ of time-inversion on
Eq.~\refg{eq:110} we receive
\begin{equation}
  \label{eq:113}
\begin{array}{l}
  Kw_{i,m}(\bm r - \bm R - \bm\rho_i, t) = \\\\\displaystyle\frac{1}{\sqrt{N}}
  \sum^{BZ}_{\bm
    k}\sum_{q = 1}^{\mu}e^{i\bm k (\bm R + \bm\rho_i)}g^*_{iq}(\bm k
  )K\varphi_{\bm k,q,m}(\bm r, t).
\end{array}  
\end{equation}

Eq.~\refg{eq:102}, on the other hand, may be written as
\begin{equation}
  \label{eq:115}
\begin{array}{l}
  Kw_{i,m}(\bm r - \bm R - \bm\rho_i, t) =  
\\\\\displaystyle\frac{1}{\sqrt{N}}
  \sum^{BZ}_{\bm
    k}\sum_{q = 1}^{\mu}e^{-i\bm k (\bm R + \bm\rho_i)}g_{iq}(\bm k
  )\nu (m)\varphi_{\bm k,q,-m}(\bm r, t)
\end{array}
\end{equation}
or, by replacing under the sum $\bm k$ by $-\bm k$, 
\begin{equation}
  \label{eq:121}
\begin{array}{l}
  Kw_{i,m}(\bm r - \bm R - \bm\rho_i, t) =  
\\\\ = \displaystyle\frac{1}{\sqrt{N}}
  \sum^{BZ}_{\bm
    k}\sum_{q = 1}^{\mu}e^{i\bm k (\bm R + \bm\rho_i)}g_{iq}(-\bm k
  )\nu (m)\varphi_{-\bm k,q,-m}(\bm r, t),
\end{array}
\end{equation}
where
\begin{equation}
  \label{eq:119}
  \nu\big(\pm\frac{1}{2}\big) = \pm 1.
\end{equation}

Comparing Eq.~\refg{eq:121} with Eq.~\refg{eq:113} we receive the two equations
\begin{equation}
  \label{eq:122}
  g^*_{iq}(\bm k ) = g_{iq}(-\bm k )
\end{equation}
and
\begin{equation}
  \label{eq:123}
  K\varphi_{\bm k,q,m}(\bm r, t) = \nu (m)\varphi_{-\bm k,q,-m}(\bm r, t).
\end{equation}
While the first Eq.~\refg{eq:122} is relatively meaningless, from the
second Eq.~\refg{eq:123} we may derive the important Eq.~\refg{eq:124}:

Eq.~\refg{eq:62} yields the two equations
\begin{equation}
  \label{eq:152}
  \varphi_{-\bm k,q,-m}(\bm r,t) = \sum_{s = -\frac{1}{2}}^{+\frac{1}{2}}
f_{-m,-s}(q,-\bm k)\psi_{-\bm k,q,-s}(\bm r, t)
\end{equation}
where now the sum runs over $-s$, and 
\begin{equation}
  \label{eq:153}
  K\varphi_{\bm k,q,m}(\bm r,t) = \sum_{s = -\frac{1}{2}}^{+\frac{1}{2}}
f^*_{ms}(q,\bm k)\nu (s)\psi_{-\bm k,q,-s}(\bm r, t)
\end{equation}
because~\cite{bc}
\begin{equation}
  \label{eq:154}
  K\psi_{\bm k,q,s}(\bm r, t) = \nu (s)\psi_{-\bm k,q,-s}(\bm r, t).
\end{equation}
\begin{thm}
\label{theorem12}
Substituting Eqs.~\refg{eq:152} and~\refg{eq:153} in Eq.~\refg{eq:123}
we obtain the fundamental condition
\begin{equation}
  \label{eq:124}
f_{ms}^*(q, -\bm k) = \pm f_{-m,-s}(q, \bm k),
\end{equation}
where the plus sign holds for $m = s$ and the minus for $m = -s$.

Within the nonadiabatic Heisenberg model, the validity of this
condition is the cause of the formation of symmetrized
Cooper pairs in superconducting bands~\cite{es1,es,josn}.
\end{thm}

This Eq.~\refg{eq:124} may evidently be written in the more compact form
\begin{equation}
  \label{eq:149}
    {\bf f}^*(q, -\bm k) = {\bf nf}(q, \bm k){\bf n}^{-1}
\end{equation}
where ${\bf n}$ is given in Eq.~\refg{eq:104}.

\subsection{$\bm k$-dependence of the matrices ${\bf f}(q, \bm k)$}
\label{sec:fqk}
Only those bands are of physical relevance in the theory of
superconductivity which are closed not before the spin-dependent
perturbation $\mathcal{H}_s$ is activated. In this section we derive
the essential property of such bands and shall give the
result in Theorem~\ref{theorem7}.

Let be $\bm k$ a point lying on the surface of the first domain in the
Brillouin zone for the space group $H$ and let be $H_{\bm k}$ the
little group of $\bm k$. In this section, $\bm k$ need not be a point
of symmetry [according to Definition~\ref{def:19}] but also may lie in
a line or a plane of symmetry. However, we only consider wave vectors
$\bm k$ at which Eq.~\refg{eq:56} is valid. Hence, in general, the
Bloch functions $\varphi_{\bm k,q}(\bm r)$ are basis functions for a
{\em one}-dimensional (single-valued) representation of $H_{\bm
  k}$. Nevertheless, in very rare cases, the Bloch function
$\varphi_{\bm k,q}(\bm r)$ can be a basis function for a degenerate
(single-valued) representation. Both cases shall be examined
separately.

Just as in Eq.~(3.1) of Ref.~\cite{ew1} we arrange the $2\mu$ Bloch spinors
$u_s(t)\varphi_{\bm k,q}(\bm r)$ in Eq.~\refg{eq:56} as column vector
\begin{equation}
  \label{eq:127}
  \Phi_{\bm k}(\bm r, t) = \left(
\begin{array}{c}
u_{+\frac{1}{2}}(t)\varphi_{\bm k,\mu}(\bm r)\\
u_{-\frac{1}{2}}(t)\varphi_{\bm k,\mu}(\bm r)\\
\vdots\\
u_{+\frac{1}{2}}(t)\varphi_{\bm k,2}(\bm r)\\
u_{-\frac{1}{2}}(t)\varphi_{\bm k,2}(\bm r)\\
u_{+\frac{1}{2}}(t)\varphi_{\bm k,1}(\bm r)\\
u_{-\frac{1}{2}}(t)\varphi_{\bm k,1}(\bm r)
\end{array}
\right)
\end{equation}
with increasing energy,
\begin{equation}
  \label{eq:150}
  E_{\bm k, q - 1} \leq E_{\bm k, q} \leq E_{\bm k, q + 1}. 
\end{equation}

Then the analogous column vector $\widetilde\Phi_{\bm k}(\bm r, t)$
consisting of the Bloch spinors $\widetilde\varphi_{\bm k,i,m}(\bm r,
t)$ in Eq.~\refg{eq:67} may be written as
\begin{equation}
  \label{eq:128}
  \widetilde\Phi_{\bm k}(\bm r, t) =
  {\bf g}^d(\bm k)\cdot {\bf f}^d(\bm k)\cdot\Phi_{\bm k}(\bm r, t)
\end{equation}
where
\begin{equation}
  \label{eq:129}
  {\bf g}^d(\bm k) = {\bf g}(\bm k) \times 
  \left( 
  \begin{array}{cc}
  1 & 0\\
  0 & 1
  \end{array}
  \right)
\end{equation}
and
\begin{equation}
  \label{eq:130}
  {\bf f}^d(\bm k) =
\left(
\begin{array}{cccc}  
  {\bf f}(\mu , \bm k) & \bm 0 & \bm 0 & \bm 0\\
  \vdots & \ddots & \vdots & \vdots \\
  \bm 0 & \bm 0 & {\bf f}(2, \bm k) & \bm 0\\
  \bm 0 & \bm 0 & \bm 0 & {\bf f}(1, \bm k)
\end{array}
\right).
\end{equation}
The matrices ${\bf g}(\bm k)$ and ${\bf f}(q, \bm k)$ are
defined by Eqs.~\refg{eq:148} and~\refg{eq:94} and still follow
Eqs.~\refg{eq:5} and~\refg{eq:86}, respectively, and
\begin{equation}
  \label{eq:135}
  \bm 0 =
  \left(
  \begin{array}{cc}
  0 & 0\\
  0 & 0
  \end{array}
  \right).
\end{equation}

The matrices ${\bf g}^d(\bm k)\cdot {\bf f}^d(\bm k)$ must satisfy
Eqs.~(4.8) and~(4.29) of Ref.~\cite{ew1} in order that the Wannier
functions are symmetry-adapted and best localized. [We shall consider
only Eq.~(4.29) of Ref.~\cite{ew1} because this equation comprises
Eq.~(4.8) {\em ibidem}].

Using the notations of the present paper, Eq.~(4.29) of
Ref.~\cite{ew1} may be written as
\begin{eqnarray}
  \label{eq:131}
  {\bf D}^{d}_{\bm k}(a)& = & \big({\bf g}^{d*}(\bm
k)\cdot {\bf f}^{d*}(\bm k)\big)^{-1}\cdot
    \overline{\widehat{\bf D}}^{d}_{\bm k}(a)\cdot\big({\bf g}^{d*}(\bm
k)\cdot {\bf f}^{d*}(\bm k)\big)\nonumber\\
&& \text{for } a \in H^d_{\bm k},
\end{eqnarray}
where the matrices ${\bf D}^{d}_{\bm k}(a)$ and
$\overline{\widehat{\bf D}}^{d}_{\bm k}(a)$ denote the representatives
of the the representations $\bm D^d_{\bm k}$ and
$\overline{\widehat{\bm D}}^{d}_{\bm k}$ given in Eqs.~\refg{eq:157}
and~\refg{eq:143}, respectively. Assume that the representations $\bm
D^d_{\bm k}$ are determined according to Theorem~\ref{theorem5}.  Then
the representations $\bm D^{\text{aff}}_{\bm k}$ and
$\overline{\widehat{\bm D}}_{\bm k}$ as well as the representations
$\bm D^d_{\bm k}$ and $\overline{\widehat{\bm D}}^d_{\bm k}$ are
equivalent for the points $\bm k$ of symmetry.  Consequently, these
representations are even equivalent in any point $\bm k$ of the
Brillouin zone because the compatibility relations are valid in a
closed band~\cite{ew1}. First, from the equivalence of $\bm
D^{\text{aff}}_{\bm k}$ and $\overline{\widehat{\bm D}}_{\bm k}$ it
follows that the equation
\begin{equation}
  \label{eq:151}
  {\bf D}^{\text{aff}}_{\bm k}(a) = {\bf g}^{*-1}(\bm
    k)\cdot
    \overline{\widehat{\bf D}}_{\bm k}(a)\cdot {\bf g}^{*}(\bm
    k)\text{ for } a \in H_{\bm k}
\end{equation}
is solvable for any $\bm k$.

\subsubsection{The Bloch functions $\varphi_{\bm
  k,q}(\bm r)$ are basis functions for a non-degenerate representation}
\label{sec:fkqnond}
In this subsection we assume that the Bloch states $\varphi_{\bm
  k,q}(\bm r)$ are basis functions for a {\em one}-dimensional
(single-valued) representation of $H_{\bm k}$.

The representations $\bm D^{d}_{\bm k}$ are the direct sum over the
double-valued representations of the Bloch spinors in the considered
band, as arranged in the column vector given in
Eq.~\refg{eq:127}. Hence, the matrices ${\bf D}^{d}_{\bm k}(a)$ on the
left hand side of Eq.~\refg{eq:131} may be written as
\begin{equation}
  \label{eq:132}
  {\bf D}^{d}_{\bm k}(a) = 
\left(
\begin{array}{cccc}  
  {\bf d}_{\bm k, \mu}(a) & 0 & 0 & 0\\
  \vdots & \ddots & \vdots & \vdots \\
  0 & 0 & {\bf d}_{\bm k, 2}(a) & 0\\
  0 & 0 & 0 & {\bf d}_{\bm k, 1}(a)
\end{array}
\right) \times{\bf d}_{1/2}(\alpha )
\end{equation}
(for $a = \{\alpha |\bm t_{\alpha}\} \in H^d_{\bm k}$), where the
Bloch state $\varphi_{\bm k,q}(\bm r)$ is basis function for the
single-valued representations $\bm d_{\bm k, q}$.

The matrices on the right hand side of Eq.~\refg{eq:131} may be
written as
\begin{equation}
  \label{eq:136}
  \begin{array}{c}
    \big({\bf g}^{d*}(\bm
    k)\cdot {\bf f}^{d*}(\bm k)\big)^{-1}\cdot
    \overline{\widehat{\bf D}}^{d}_{\bm k}(a)\cdot\big({\bf g}^{d*}(\bm
    k)\cdot {\bf f}^{d*}(\bm k)\big) =\\
    \\
    \big({\bf f}^{d*}(\bm k)\big)^{-1}\cdot\Big[\big({\bf g}^{d*}(\bm
    k)\big)^{-1}\cdot
    \overline{\widehat{\bf D}}^{d}_{\bm k}(a)\cdot {\bf g}^{d*}(\bm
    k)\Big]\cdot {\bf f}^{d*}(\bm k).
  \end{array}
\end{equation}
Using Eqs.~\refg{eq:129},~\refg{eq:142} and~\refg{eq:151} we may write
the matrices between the square brackets as
\begin{equation}
  \label{eq:137}
\begin{array}{ll}
  \big({\bf g}^{d*}(\bm
    k)\big)^{-1}\cdot
    \overline{\widehat{\bf D}}^{d}_{\bm k}(a)\cdot {\bf g}^{d*}(\bm
    k) &=\\
\\
  \Big({\bf g}^{*-1}(\bm
    k)\cdot
    \overline{\widehat{\bf D}}_{\bm k}(a)\cdot {\bf g}^{*}(\bm
    k)\Big) \times {\bf d}_{1/2}(\alpha ) & =\\
\\
  {\bf D}^{\text{aff}}_{\bm k}(a) \times {\bf d}_{1/2}(\alpha ) & =\\
\\
\multicolumn{2}{l}{
\left(
\begin{array}{cccc}  
  {\bf d}_{\bm k, \mu}^{\text{aff}}(a) & 0 & 0 & 0\\
  \vdots & \ddots & \vdots & \vdots \\
  0 & 0 & {\bf d}_{\bm k, 2}^{\text{aff}}(a) & 0\\
  0 & 0 & 0 & {\bf d}_{\bm k, 1}^{\text{aff}}(a)
\end{array}
\right) \times{\bf d}_{1/2}(\alpha ),
}
\end{array}
\end{equation}
where again the matrices ${\bf d}_{\bm k, q}^{\text{aff}}(a)$ form
single-valued one-dimensional representations ${\bm d}_{\bm k,
  q}^{\text{aff}}$. Remember that [Definition~\ref{def:20}] the
single-valued representations $\bm d_{\bm k, q}^{\text{aff}}$ are not
associated to the Bloch functions of the considered band but are
fixed by the representation $\bm D$ defining the spin-dependent
Wannier functions.

Eq.~\refg{eq:137} shows that also the matrices between the square
brackets form a representation being the direct sum over double-valued
representations and, hence, Eq.~\refg{eq:131} splits into the $\mu$
equations
\begin{equation}
  \label{eq:140}
  \bm d_{\bm k, q}\otimes\bm d_{1/2} = {\bf f}^{*-1}(q, \bm k)
  \cdot \big(\bm d_{\bm k, q}^{\text{aff}}\otimes\bm d_{1/2}\big)\cdot {\bf f}^*(q, \bm k),
\end{equation}
($1 \leq q \leq \mu$), which are solvable because the representations
$\bm D^d_{\bm k}$ and $\overline{\widehat{\bm D}}^d_{\bm k}$ and,
hence, also the representations $\bm d_{\bm k, q}\otimes\bm d_{1/2}$
and $\bm d_{\bm k, q}^{\text{aff}}\otimes\bm d_{1/2}$ are equivalent.

We now distinguish between two possibilities:
\begin{itemize}
\item If the considered energy band was already closed before the
  spin-dependent perturbation $\mathcal{H}_s$ was activated, then the
  affiliated single-valued band actually exists as closed band in the
  band structure of the material under consideration and, thus, the
  representations $\bm d_{\bm k, q}$ and $\bm d_{\bm k,
    q}^{\text{aff}}$ are equal,
  \begin{equation}
    \label{eq:138}
    \bm d_{\bm k, q} = \bm d_{\bm k, q}^{\text{aff}}.
  \end{equation}
  Hence, all the $\mu$ equations~\refg{eq:140} are solved by
\begin{equation}
  \label{eq:139}
{\bf f}(q, \bm k) \equiv {\bf 1},  
\end{equation}
with the consequence that the Wannier functions are, in fact, not
spin-dependent but are usual Wannier functions as defined in
Eq.~\refg{eq:1}.
\item If the considered energy band was not closed before the
  spin-dependent perturbation $\mathcal{H}_s$ was activated, then not
  all the representations $\bm d_{\bm k, q}$ are equal to $\bm d_{\bm
    k, q}^{\text{aff}}$. Evidently, the $q$th equation is {\em not}
  solved by ${\bf f}(q, \bm k) \equiv \bm 1$ when $\bm d_{\bm k, q}
  \neq \bm d_{\bm k, q}^{\text{aff}}$ and, consequently, the Wannier
  function actually are spin-dependent.
\end{itemize}

We summarize this result in Theorem~\ref{theorem7}.
\begin{thm}
\label{theorem7}
If the considered energy band was not closed before the spin-dependent
perturbation $\mathcal{H}_s$ was activated, the matrices ${\bf f}(q,
\bm k)$ in Eq.~\refg{eq:62} cannot be chosen independent of $\bm k$.
\end{thm}  

In the Sec.~\ref{sec:niob} the matrix ${\bf f}(q, \bm k)$
shall by determined for some points in the Brillouin zone of niobium.

\subsubsection{The Bloch functions $\varphi_{\bm
  k,q}(\bm r)$ are basis functions for a degenerate representation}
\label{sec:fkqd}
In rare cases, it can happen that at a special point $\bm k$ some of
the Bloch states $\varphi_{\bm k,q}(\bm r)$ are basis functions for a
degenerate (single-valued) representation and that this degeneracy is
{\em not} removed by the perturbation $\mathcal{H}_s$. For example,
each of the two superconducting bands in the space group $P4/nmm =
\Gamma_q D^7_{4h}$ (129) listed in Table 3 (b) of Ref.~\cite{lafeaso2}
consist of two branches degenerate at points $M$ and $A$. The
single-valued Bands 1 and 2 in Table 3 (a) of Ref.~\cite{lafeaso2} are
affiliated to the superconducting Band 1 in Table 3 (b) {\em ibidem};
Bands 3 and 4 in Table 3 (a) are affiliated to Band 2 in Table 3 (b).

It is crucial for the localization of the spin-dependent Wannier
functions that also in this case Eq.~\refg{eq:131} is solvable.
We reveal the solubility of this equation on the example of the
bands listed in Table 3 of Ref.~\cite{lafeaso2}.

At point $M$ in each of these bands, Eq.~\refg{eq:140} may be written
as
\begin{equation}
  \label{eq:118}
  \bm d_{\bm k_M}\otimes\bm d_{1/2} = \big({\bf f}^{d*}(\bm k_M)\big)^{-1}
  \cdot \big(\bm d_{\bm k_M}^{\text{aff}}\otimes\bm d_{1/2}\big)\cdot {\bf f}^{d*}(\bm k_M),
\end{equation}
where $\bm d_{\bm k_M}$ and $\bm d_{\bm k_M}^{\text{aff}}$ now are
{\em two}-dimensional (single-valued) representations and the matrix
${\bf f}^{d}(\bm k_M)$ now is four-dimensional,
\begin{equation}
  \label{eq:155}
  {\bf f}^d(\bm k_M) =
\left(
\begin{array}{cc}  
{\bf f}(2, \bm k_M) & \bm 0\\
 \bm 0 & {\bf f}(1, \bm k_M)
\end{array}
\right),
\end{equation}
see Eq.~\refg{eq:130}.

Though $\bm d_{\bm k_M}\otimes\bm d_{1/2}$ and $\bm d_{\bm
  k_M}^{\text{aff}}\otimes\bm d_{1/2}$ again are equivalent, it is not
immediately evident that Eq.~\refg{eq:118} is solvable because ${\bf
  f}^{d}(\bm k_M)$ is not a general $4\times 4$ matrix. However, also
the representations $\bm d_{\bm k_M}\otimes\bm d_{1/2}$ and $\bm
d_{\bm k_M}^{\text{aff}}\otimes\bm d_{1/2}$ have a very special form
since they may be written simply as Kronecker products.
Eq.~\refg{eq:118} indeed is solvable since it expresses the most
general unitary transformation between these special representations.

For instance, consider the point $M$ of one of the bands in Table 3
(b) of Ref.~\cite{lafeaso2} and let be $\bm d_{\bm k_M} = M_3$ given
by the calculated band structure of the material under
consideration. In addition, let us choose Band 1 in Table 3 (a) of
Ref.~\cite{lafeaso2} as affiliated single-valued band. Thus, we have
$\bm d_{\bm k_M}^{\text{aff}} = M_2$ and Eq.~\refg{eq:118} is solved
by
\begin{equation}
  \label{eq:156}
  {\bf f}^d(\bm k_M) =
\left(
\begin{array}{cc}  
\left(
\begin{array}{cc}
0 & -i\\
1 & 0
\end{array}
\right)
& \bm 0\\
 \bm 0 & 
\left(
\begin{array}{cc}
0 & 1\\
-i & 0
\end{array}
\right)
\end{array}
\right),
\end{equation}
as it may be determined by means of the tables given in Ref.~\cite{bc}. 

Though both Band 1 and Band 2 in Table 3 (b) of Ref.~\cite{lafeaso2}
are mathematically correct superconducting bands, they cannot be
occupied in undoped LaFeAsO~\cite{lafeaso2} which, consequently, is
not superconducting.

\subsubsection{Additions}
\label{sec:additions}

In this subsection we show that neither Eq.~\refg{eq:122} nor
Eq.~\refg{eq:149} is inconsistent with Eq.~\refg{eq:131}. Remember
that in this section we only consider points $\bm k$ at which 
Eq.~\refg{eq:56} is valid.

First, taking the complex conjugate of Eq.~\refg{eq:151}, we receive
with $\bm D^{\text{aff}*}_{\bm k} = \bm D^{\text{aff}}_{-\bm k}$ and
$\overline{\widehat{\bm D}}^*_{\bm k} = \overline{\widehat{\bm
    D}}_{-\bm k}$ the condition
\begin{equation}
  \label{eq:146}
  \bm D^{\text{aff}}_{-\bm k} = {\bf g}^{-1}(\bm k)\cdot
  \overline{\widehat{\bm D}}_{-\bm k}\cdot {\bf g}(\bm k) 
\end{equation}
showing that we may chose  
\begin{equation}
  \label{eq:120}
  {\bf g}^*(-\bm k) = {\bf g}(\bm k)
\end{equation}
and, hence, Eq.~\refg{eq:122} is consistent with Eq.~\refg{eq:146} and,
consequently, with Eq.\refg{eq:131}.

Secondly, transforming the complex conjugate of Eq.~\refg{eq:140} with
the matrix ${\bf n}$ in Eq.~\refg{eq:104} and using
$\bm d^*_{\bm k, q} = \bm d_{-\bm k, q}$, $\bm
d_{\bm k, q}^{\text{aff}*} = \bm d_{- \bm k, q}^{\text{aff}}$, and
\begin{equation}
  \label{eq:147}
      \bm d^*_{1/2} = {\bf n}^{-1}\bm d_{1/2}{\bf n} 
\end{equation}
(see Eq.~\refg{eq:114}), we obtain the equation
\begin{equation}
  \label{eq:125}
    \begin{array}{l}
      \bm d_{-\bm k, q}\otimes\bm d_{1/2} = \\\big({\bf nf}(q, 
      \bm k){\bf n}^{-1}\big)^{-1}
      \cdot (\bm d_{-\bm k, q}^{\text{aff}}\otimes\bm d_{1/2})\cdot\big({\bf nf}(q, \bm k)
      {\bf n}^{-1}\big)
    \end{array}
\end{equation}
showing that, in fact, we may chose
  $${\bf f}^*(q, -\bm k) = {\bf nf}(q, \bm k){\bf n}^{-1}.$$
Hence, Eq.~\refg{eq:149} is consistent with
Eq.~\refg{eq:131}.

\subsection{Example: Band structure of Niobium}
\label{sec:niob}

\begin{figure}  
\centering
\resizebox{0.5\textwidth}{!}{%
\includegraphics{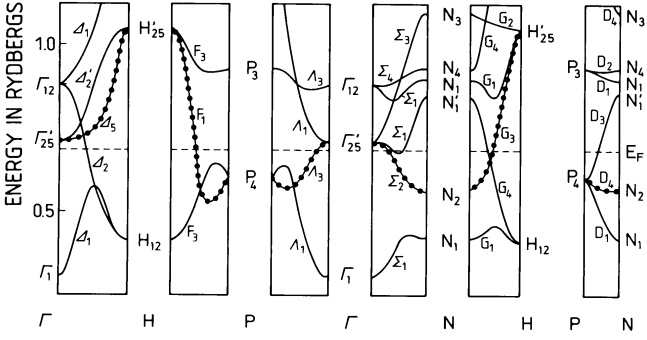}}
\caption{Band structure of Nb after Mattheis~\protect\cite{mattheis}. The dotted line denotes the superconducting band.}
\label{figure1}
\end{figure}  

Consider the superconducting band [Definition~\ref{def:22}] of niobium
in Fig.~\ref{figure1}, as denoted by the dotted line. At the four
points of symmetry $\Gamma$, $H$, $N$, and $P$ in the Brillouin zone
for the space group $O^9_h$ of niobium, this band is characterized by
the representations
$$
\Gamma_{25}',\ H_{25}',\ N_{2}, \text{ and }P_{4}  
$$
of $O^9_h$ in the familiar notation of Bouckaert, Smoluchowski and
Wigner~\cite{bouckaert}, which may be written as
\begin{equation}
  \label{eq:89}
\Gamma^+_{5},\ H^+_{5},\ N^+_{4}, \text{ and }P_{5},  
\end{equation}
respectively, in the notation of Bradley and Cracknell ~\cite{bc} (see
Tables 5.7 and 5.8 {\em ibidem}) which is consistently used in our
papers.  When we take into account that the electrons possess a spin,
we receive
\begin{equation}
  \label{eq:90}
\begin{array}{lllllll}
\Gamma^+_{5} & \otimes & d_{1/2} &=& \Gamma_7^+ & \oplus\ & \Gamma_8^+,
\nonumber\\ 
H^+_{5} & \otimes & d_{1/2} &=& H_7^+ & \oplus\ & H_8^+,
\nonumber\\ 
P_5 & \otimes & d_{1/2} &=& P_7 & \oplus\ & P_8,
\nonumber\\ 
N^+_4 & \otimes & d_{1/2} &=& N_5^+.
\end{array}
\end{equation}

Hence, at the points $\Gamma$, $H$, $P$, and $N$ the Bloch spinors can
be transformed in such a way that at each of the four points $\Gamma$,
$H$, $N$, and $P$ two spinors form basis functions for the
double-valued representations
\begin{equation}
  \label{eq:91}
  \Gamma_7^+,\ H_7^+,\ P_7, \text{ and } N_5^+,
\end{equation}
respectively. We may unitarily transform the Bloch spinors $\psi_{\bm
  k,q,s}(\bm r, t)$ of this single energy band characterized by the
representations~\refg{eq:91} into best localized and symmetry-adapted
spin-dependent Wannier functions because Theorem~\ref{theorem5}
yields with $H_0 = O_h$, $\mu = 1$, $\bm \rho_1 = \bm 0$, $G_{0p} =
H_0 = O_h$, and $\bm d_1 = \Gamma^+_2$ first the single-valued
representations
\begin{equation}
  \label{eq:97}
  \bm D^{\text{aff}}_{\Gamma} = \Gamma^+_2,\ 
  \bm D^{\text{aff}}_{H} = H^+_2,\ 
  \bm D^{\text{aff}}_{P} = P_2,\ \text{and}\ 
  \bm D^{\text{aff}}_{N} = N^+_3
\end{equation}
and then, with Eq.~\refg{eq:157}, the double-valued
representations~\refg{eq:91}.

The representations in Eq.~\refg{eq:97} define (the only)
single-valued band affiliated to the superconducting band defined by
the representations in Eq.~\refg{eq:91} [Definition~\ref{def:20}].
The representation $\bm D$ defining the spin-dependent Wannier
functions [Definition~\ref{def:17}] is equal to $\Gamma^+_2$,
\begin{equation}
  \label{eq:3}
  \bm D = \Gamma^+_2.
\end{equation}
$\bm D$ is one-dimensional since we have one Nb atom in the unit
cell. The spin-dependent Wannier functions may be chosen
symmetry-adapted to the magnetic group in Eq.~\refg{eq:100} because
$\Gamma^+_2$ is real. 

The Bloch functions of the superconducting band {\em cannot} be
unitarily transformed into usual Wannier functions which are best
localized and symmetry-adapted to $O^9_h$ since it was not closed
before the spin-dependent perturbation $\mathcal{H}_s$ was activated.
Thus [Theorem~\ref{theorem7}], we cannot choose the matrix ${\bf f}(1,
\bm k)$ in Eq.~\refg{eq:62} (with $q = 1$ since we only have one
branch in the superconducting band of Nb) independent of $\bm k$ when
we demand that the Wannier functions are best localized and
symmetry-adapted. This important statement shall be demonstrated by an
example:

Consider the point $N$ with the wave vector $\bm k_N$ in the first
domain of the Brillouin zone for $O^9_h$.  The representations $\bm d_{\bm
  k_N, 1}^{\text{aff}}$ and $\bm d_{\bm k_N, 1}$ in Eq.~\refg{eq:140} are
given by Eqs.~\refg{eq:97} and~\refg{eq:89},
\begin{equation}
  \label{eq:82}
  \bm d_{\bm k_N, 1}^{\text{aff}} = N^+_3
\end{equation}
and
\begin{equation}
  \label{eq:84}
  \bm d_{\bm k_N, 1} = N^+_4.
\end{equation}
Thus, Eq.~\refg{eq:140} may be written as
\begin{equation}
  \label{eq:83}
  N^+_4\otimes\bm d_{1/2} = \\
  ({\bf f}^{*}(1, \bm k_N))^{-1}\cdot (N^+_3\otimes\bm d_{1/2})
  \cdot {\bf f}^*(1, \bm k_N).
\end{equation}
This equation is solvable since both representations $N^+_4\otimes\bm
d_{1/2}$ and $N^+_3\otimes\bm d_{1/2}$ are equivalent, but it is
evidently not solved by ${\bf f}(1, \bm k_N) = {\bf 1}$. In fact, we
receive 
\begin{equation}
  \label{eq:144}
  {\bf f}(1, \bm k_N) = 
  \left(
  \begin{array}{cc}
  0 & 1\\
  -i & 0
  \end{array}
  \right)
\end{equation}
by means of Tables 5.7 and 6.1 of Ref.~\cite{bc}.  This is the value
of ${\bf f}(1, \bm k)$ also on the planes of symmetry intersecting at
$N$ in the neighborhood of $N$. Further away from $N$, however, ${\bf
  f}(1, \bm k)$ may change since it is $\bm k$ dependent.

In the same way, we find
\begin{equation}
  \label{eq:145}
  {\bf f}(1, \bm k_F) = \frac{1}{\sqrt{3}}
  \left(
  \begin{array}{cc}
  -i & -1 + i\\
  1 + i & i
  \end{array}
  \right)
  \end{equation}
for the points $\bm k_F$ on the line $F$. 

Eqs.~\refg{eq:144} and~\refg{eq:145} demonstrate that ${\bf f}(1, \bm
k)$ cannot be chosen independent of $\bm k$ in the superconducting
band of niobium.      

\setcounter{equation}{0}
\setcounter{definition}{0}
\setcounter{thm}{0}
\section{Conclusion}
\label{sec:summary}
In the present paper we gave the group theory of best localized and
symmetry-adapted Wannier functions with the expectation that it will
be helpful to determine the symmetry of the Wannier functions in the
band structure of any given material. The paper is written in such a
way that it should be possible to create a computer program automating
the determination of Wannier functions.

In this paper we restricted ourselves to Wannier functions that define
magnetic or superconducting bands. That means that we only considered
Wannier functions centered at the atomic positions. When other
physical phenomena shall be explored, as, e.g., the metallic bound,
other Wannier functions may be needed which are centered at other
positions, e.g., between the atoms. It should be noted that
Refs.~\cite{ew1},~\cite{ew2} and~\cite{ew3} define best localized and
symmetry-adapted Wannier functions in general terms which may be
centered at a variety of positions $\bm \rho_i$ being different from
the positions of the atoms.

\acknowledgements 
We are indebted to Guido Schmitz for his support of our work. 

%


\end{document}